\documentclass[12pt,a4paper]{article}

\usepackage[T1]{fontenc}
\usepackage[utf8]{inputenc}
\usepackage[english]{babel}

\usepackage{mathrsfs}

\usepackage{array}
\usepackage{amsmath}
\usepackage{amssymb}
\usepackage{epsf}

\usepackage{amsthm}
\usepackage{ulem}

\usepackage{slashed}
\usepackage{geometry}

\usepackage[all]{xy}
\usepackage{feynmf}
\usepackage{graphicx}

\usepackage{amsfonts}

\def\ben{\begin{equation}}
\def\een{\end{equation}}
\def\bena{\begin{eqnarray}}
\def\eena{\end{eqnarray}}

\theoremstyle{definition}
\newtheorem{thm}{Theorem}

\newtheorem{lemma}{Lemma}[section]
\newtheorem{prop}{Proposition}[section]

\newcommand{\al}{\alpha}
\newcommand{\be}{\beta}
\newcommand{\de}{\delta}

\newcommand{\vep}{\varepsilon}
\newcommand{\ga}{\gamma}
\newcommand{\Ga}{\Gamma}

\newcommand{\la}{\lambda}
\newcommand{\si}{\sigma}

\newcommand{\vp}{\varphi}
\newcommand{\La}{\Lambda}

\newcommand{\Lao}{\Lambda_0}

\newcommand{\oc}{\bar{c}}
\newcommand{\opsi}{\bar{\psi}}
\newcommand{\ochi}{\bar{\chi}}
\newcommand{\oeta}{\bar{\eta}}
\newcommand{\pa}{\partial}

\newcommand{\ti}[1]{\tilde{#1}}

\newcommand{\eq}{\begin{equation}}
\newcommand{\eqe}{\end{equation}}

\newcounter{saveeqn}
\newcommand{\alpheqn}{\setcounter{saveeqn}{\value{equation}}
\setcounter{equation}{0}
\addtocounter{saveeqn}{1}
\renewcommand{\theequation}{\mbox{\arabic{saveeqn}\alph{equation}}}}
\newcommand{\reseteqn}{\setcounter{equation}{\value{saveeqn}}%
\renewcommand{\theequation}{\arabic{equation}}}

\newcommand{\jn}{j_{\nu}}

\newcommand{\ca}{c}

\newcommand{\cb}{{\bar c}}

\newcommand{\An}{A_{\nu}}
\newcommand{\et}{\eta}

\newcommand{\etb}{{\bar \eta}}

\newcommand{\LLz}{\Lambda, \Lambda_0}
\newcommand{\LLzz}{\Lambda_0, \Lambda_0}
\newcommand{\Lz}{\,0, \Lambda_0}

\newcommand{\uca}{\underline c}
\newcommand{\ucb}{\underline {\bar{c}}}
\newcommand{\uAn}{\underline A_{\nu}}
\newcommand{\uFi}{\underline \Phi }

\newcommand{\mr}{{\mathbb R}}

\begin{document}


\title{Regularized path integrals and anomalies\\
 - U(1) chiral gauge theory -}
\author{Christoph
  Kopper\footnote{christoph.kopper@cpht.polytechnique.fr} ,\hspace{1cm}
Benjamin L\'ev\^eque\footnote{benjamin.leveque@cpht.polytechnique.fr} \\
Centre de Physique Th{\'e}orique, CNRS, UMR 7644\\
Ecole Polytechnique,
F-91128 Palaiseau, France
}

\date{24 October 2011}

\maketitle
\begin{abstract}
We analyse the origin of the Adler-Bell-Jackiw (ABJ) anomaly of chiral U(1)
gauge theory within the framework of regularized path
integrals. Momentum or position space regulators allow for 
mathematically well-defined path integrals but violate local gauge
symmetry. It is known how (nonanomalous) gauge symmetry can be recovered 
in the renorma\-lized theory in this case \cite{KM}. 
Here we analyse U(1) chiral gauge theory
to show how the appearance of anomalies manifests itself 
in such a context. We show that the
three-photon amplitude leads to a violation 
of  the Slavnov-Taylor-Identities which  
cannot be restored on taking the UV limit in the renormalized theory.  
We point out that this fact is related to the  nonanalyticity of this
amplitude in the infrared region.

\end{abstract}

\tableofcontents

\section{Introduction}

When analysing a quantum field theory model one typically starts
from a lagrangian encoding its field content and symmetries. Still,
writing a lagrangian generally does not define the theory, not
even when restricting to perturbation theory. This is due in 
particular to the need of renormalization which requires to modify
the lagrangian by adding counter terms, or, in the language of
the Wilson renormalization group  \cite{Wi}, to follow the flow of the 
relevant parameters of the theory. It may then turn out that the
process of renormalization does not fully respect the symmetry
structure of the initial lagrangian. If a symmetry is inevitably
broken by the quantum corrections, 
one talks of an anomalous symmetry. 
It may also happen that the symmetry is only broken at an intermediate
stage through regulators which make the theory well-defined and can be
recovered, once these regulators are taken away again.
It is generally admitted that theories  which can be 
fully regularized
without breaking any of their symmetries, cannot   
be anomalous.  

In  this paper we want to come back to the chiral U(1) gauge theory 
- which is known  to be anomalous \cite{Ad1}, \cite{BJ}, 
\cite{FS}, \cite{Bert},
\cite{Zi2} - 
in a momentum space regularization scheme, 
which breaks gauge invariance from the beginning.
Such regularizations are used when establishing 
the differential flow equations 
\cite{WH} of the renormalization group  \cite{Wi},
which allow for an elegant inductive approach
to  perturbative renormalization theory \cite{Po}.

Most often perturbative renormalization of gauge theories is performed
with the aid of dimensional regularization which
at first sight  respects local gauge symmetry. Most of the work 
and in particular most of the calculations have been done in 
this scheme ever since it has been known to exist. 
For chiral gauge theories containing the four-dimensional 
Levi-Civita tensor $\vep_{\mu\nu\rho\si}\,$, dimensional regularization  
does not fully respect the gauge symmetry however,
since this tensor does not have a straightforward generalization
to $4+\vep$ dimensions.    
In spite of its great advantages the dimensional scheme also has some
drawbacks, mainly on the mathematical 
side\footnote{for example when calculating the three-photon-amplitude
analysed in App. \ref{triang} in the dimensional scheme, 
it is often  stated  that this amplitude or its derivatives
are {\it arbitrary} in some
sense. Thus one may ask oneself in which sense and at which stage
the starting point is well-defined mathematically~; as it is  e.g.
in a momentum-space regularized version of the theory.}.  
It not only defies to be given rigorous meaning in path integral
formulations, it does not  even directly apply in a mathematical sense 
to perturbative Green functions as a whole without splitting 
them into graphs. Thus, in some sense it is farthest away
from nonperturbative analysis. 

On the other hand analysis of symmetries and functional 
relations in field theory are largely based on path integral
formulations.
It therefore seems to be important to study gauge theories in the
rigorous framework of regularized path integrals on which the
flow equations are based.
A proof of perturbative renormalizability of spontaneously broken
SU(2)-Yang-Mills theory with the aid of flow equation was performed
in \cite{KM}. In \cite{KKQ} an analysis of QED with massive photons
was performed. Its extension to massless photons in \cite{KKQ2} 
became technically quite involved and could (should) be improved nowadays.
A fully rigorous analysis of QCD in this framework, including the infrared 
part of the problem, still has to be performed.  
 
Let us shortly comment on the strategy of proof of \cite{KM}. 
The (ultraviolet) power  counting part of the flow equation
renormalization proof is universal and simple for all 
renormalizable theories. For gauge theories
 we have to show that gauge
invariance can be restored when the cutoffs are taken away. On the level
of the Green functions  (which are not gauge invariant) 
this  means  that we have to verify the Slavnov-Taylor identities (STI)
 of the theory. They allow to argue that 
physical quantities such as the S-matrix are gauge-invariant \cite{Zi}.
On analysing the flow equations (FE) for a gauge theory one realizes that 
the restoration of the STI depends on the choice of the
renormalization conditions chosen and is not true in general.
More precisely, since gauge invariance is violated in the regularized
theory, the renormalization group flow will generally produce
nonvanishing contributions to all those relevant parameters of the
theory, which are forbidden by gauge invariance. 
The question is then: Can we use the freedom in adjusting the
renormalization conditions such that the STI are nevertheless 
restored in the end? To answer this question a first observation
is crucial: The violation of
the STI in the regularized theory can be expressed through Green functions
carrying an operator insertion, which depends on the regulators. FE theory
for such insertions tells us that these Green functions will vanish once the
cutoffs are removed, if we achieve renormalization conditions on
the noninserted Green functions 
such that the inserted ones, which are calculated from those, 
have vanishing renormalization conditions
for all relevant terms, i.e. up to the dimension of the insertion 
(which  turns out to be  5).  In case of 
spontaneously broken Yang-Mills
theory as well as for QED it could then be shown that there exist
classes of renormalization conditions such that the relevant part 
of the STI vanishes, and in consequence such that the STI are restored
after taking away the cutoffs.

In the present paper we want to analyse the mechanism behind the
appearance of the anomaly in chiral U(1) gauge theory
in this framework. As a consequence of the previous
remarks an anomaly should manifest itself through the appearance
of a finite relevant contribution to the STI which cannot be 
eliminated by a suitable choice of renormalization conditions. 
Our analysis reveals that this  appearance is closely related
to the infrared divergences of the chiral gauge theory. In fact
it will turn out that complete Bose symmetry together with
analyticity around zero momentum - which would hold in
fully massive theories - would prevent the appearance of the 
ABJ anomaly.
The deduction of the anomalous Ward or Slavnov-Taylor identities
proceeds in the same way  as in the SU(2)-case. There is no room
for a contribution from the integration measure, which seems 
to be in contrast with the deduction of the anomaly by Fujikawa
\cite{Fu}, \cite{FuS}.
In this respect, we discuss the Jacobian of regularized 
BRS-transformations, and we also discuss  Fujikawa's argument.
We note that recently chiral anomalies have also been analysed
nonperturbatively in two-dimensional models like the Thirring
model \cite{Ma}, \cite{Fa}. Conceptually this approach is close
to ours since it is also based on reglarized path integrals,
which in this case can be analysed constructively, i.e. beyond
perturbation theory.

Our paper is organized as follows. In section \ref{class} we 
introduce the classical action of the chiral U(1) gauge
theory, its symmetries and
 the abelian BRST-transformations \cite{BRS}, \cite{Ty}.
In section \ref{rere}  we introduce 
regularized path integrals, certain  concepts from FE theory, 
and we recall the statements  on renormalizability we need. 
In particular we introduce the above mentioned operator
insertions.
When using the FE it is natural to analyse the generating functional
of free propagator amputated Schwinger functions. 
The analysis of the STI is however technically simpler for 
one-particle irreducible vertex functions so that we
introduce the generating  functionals of both,
 together with the corresponding renormalizability statements.
In section \ref{viosti} we derive the violated Slavnov-Taylor 
identities (VSTI) for the regularized theory emphasizing the 
terms related to the anomaly. Using explicit results on
the regularized three-photon-amplitude we show that {\it the 
STI cannot be restored in the UV limit for any choice
of renormalization conditions.}
As regards the general aspects of path integral analysis 
we try to keep the presentation in sections  \ref{rere}  and \ref{viosti} 
short, referring to the more detailed analogous deductions 
presented in \cite{KM} in 
the technically more involved nonabelian case.\\
In the appendices 
we analyse the ABJ anomaly in the regularized theory
and reveal its relation to the infrared singularity of the
massless fermion chiral gauge theory (App. \ref{triang});
we show that for straightforward regularizations the Jacobian 
associated with the BRS-transformation in the path integral
equals 1 (App. \ref{appa}); and  we shortly comment on 
Fujikawa's argument (App. \ref{fuji}).

\section {The classical action of chiral $U(1)$ gauge theory \label{class} }

We consider the  axial-vector-coupling abelian gauge theory with  
fermions and massive axial-vector gauge bosons,  in euclidean signature. 
We will mainly restrict to massless fermions.
The classical action has the form
\begin{equation}\label{y1}
S_{inv} = \int dx \left\{ \frac14 F^{\mu\nu} F_{\mu\nu} 
+\overline{\psi}(i \slashed{\partial} +g\slashed{A}\gamma_5)\psi\right\}
\ ,\, \ \
\mbox{with }\  \int\! dx\, \equiv \int_{\mathbb{R}^4}\! d^4x\ . 
\end{equation}
The field strength tensor is defined as  
\begin{equation}\label{y2}
F_{\mu\nu}(x) = \partial_{\mu}A_{\nu}(x) -
\partial_{\nu}A_{\mu}(x)\ . 
\end{equation}
 The coupling parameter $g\,$ is real. 
The Euclidean Dirac matrices verify the anticommutation 
relations $\left\{ \gamma_\mu,\gamma_\nu\right\}=-2\delta_{\mu\nu}\,$,
 and we adopt the  convention
\[
\gamma_5 \equiv -\gamma_0\gamma_1\gamma_2\gamma_3\ ,
\] 
such that $\gamma_5^2=1$. For massless fermions the action 
(\ref{y1}) is invariant
under local gauge transformations of the fields  
\begin{equation}
\label{y4}
A_{\mu}(x) \rightarrow A_{\mu}(x)  +  \pa_\mu u(x)
,\ \psi(x)\rightarrow e^{igu(x)\gamma_5}\psi(x)
,\ \overline{\psi}(x)\rightarrow  \overline{\psi}(x)e^{igu(x)\gamma_5}
\end{equation}
with $u: \mathbf{R}^4 \to \mathbf{R}\,$, smooth. Mass terms 
for fermions are excluded by global (chiral) gauge symmetry. 

 Aiming at a quantized theory, pure gauge degrees of
freedom have to be eliminated. We choose the standard 
covariant gauge fixing with $\alpha \in \mathbf{R}_+$, 
and we also introduce a mass $M>0$ for the gauge field,
and thus add the following contribution to the action 
\begin{equation}
S_{g.f.} = \int dx\left\{ \frac{1}{2\alpha}\, (\partial_{\mu}A_{\mu})^2 
+ \frac{M^2}{2}\, A_{\mu} A_{\mu}\right\} .
\label{y7}
\end{equation}
With regard to functional integration this condition is
implemented by introducing (bosonic but) 
anticommuting\footnote{the fields 
$\psi,\overline{\psi},c,\overline{c}$ anticommute among each other}
Faddeev-Popov ghost and
antighost fields $c$ and $\oc$ \cite{FP}, \cite{FS},
respectively, and forming with these scalar fields the
additional term in the action 
\begin{equation}\label{y8}
 S_{gh} =  \int dx\
    \oc \,(\partial_{\mu}\partial_{\mu} - \alpha M^2)
    \, c\ .
\end{equation} 
Hence, we have the total "classical action" 
\begin{equation}
S_{\rm BRS} = S_{\rm inv} + S_{\rm g.f.} + S_{\rm gh}\ ,
\label{y9a}
\end{equation}
which is decomposed as
\begin{equation}
S_{\rm BRS} = \int dx \left\{ {\cal L}_{\rm quad} (x) + {\cal L}_{\rm
int}(x) \right\}
\label{y9b}
\end{equation}    
into its quadratic part
\eq
\label{y10}  
{\cal L}_{\rm quad}\ =\ \frac14 \,
F^{\mu\nu}F_{\mu\nu}+ \frac{1}{2\alpha}\,
(\partial_{\mu}A_{\mu})^2 + \frac12\, M^2A_{\mu}A_{\mu} 
+\overline{\psi}\,i\slashed{\partial}\,\psi
          - \bar{c} (- \Delta + \alpha M^2) c \ ,
\eqe
where\footnote{we use the summation convention}
 $\Delta \equiv \partial_{\mu}
\partial_{\mu}\,$, 
and into its interaction part 
\begin{eqnarray}
{\cal L}_{\rm int} \ =\  g\, 
\overline{\psi}\,\slashed A \, \gamma_5 \,\psi\ .
\label{y11}
\end{eqnarray}

We impose the following transformation properties
of the fields under the discrete symmetries 
of charge conjugation $C\,$ and parity $P\,$~:
\begin{eqnarray*}
&&A_\mu\stackrel{C}{\rightarrow} A_\mu\,,\quad 
\psi \stackrel{C}{\rightarrow}\psi^c
=i\gamma_2\gamma_0\overline{\psi}^T,\quad 
\overline{\psi} \stackrel{C}{\rightarrow} i\psi^T \gamma_0\,\gamma_2\,, \quad
c \stackrel{C}{\rightarrow}c,\quad 
\overline{c} \stackrel{C}{\rightarrow}\overline{c}\,, \\ 
&&A_0(x)\stackrel{P}{\rightarrow} -A_0(\tilde{x}),\quad 
A_i(x)\stackrel{P}{\rightarrow} A_i(\tilde{x})\, ,\quad
\psi(x)\stackrel{P}{\rightarrow}\eta \gamma_0 \psi(\tilde{x}), \quad
\overline{\psi} \stackrel{P}{\rightarrow} \eta^*
\overline{\psi}(\tilde{x})\,\ga_0\ ,\\
&&c(x) \stackrel{P}{\rightarrow}-c(\tilde{x}),\quad 
\overline{c}(x) \stackrel{P}{\rightarrow}-\overline{c}(\tilde{x})\ .
\end{eqnarray*}
Here $\eta\,$   is an undetermined phase factor,
 and we set $\tilde{x}\equiv (x_0,-\vec{x})\,$.
Note in particular that $A_\mu\,$ transforms as an axial vector.
\\
As a prerequisite to state the symmetries of $S_{BRS}$
 (\ref{y9b}), composite classical fields are introduced as follows:
\eq
\label{BRS}
\rho_{\mu} = \partial_\mu c\ ,\quad
\rho ^{j} = ig(\gamma_5\,\psi)^j\, c\ ,\quad
\overline{\rho}^{\,j} = ig\,(\overline{\psi}\,\gamma_5)^j\,  c\ \  .  
\eqe
The classical action $S_{BRS}$ (\ref{y9b}) then  shows the
following symmetries~:
\begin{enumerate}
\item[i)] Euclidean invariance: $S_{BRS}$ is an O(4)-scalar.
\item[ii)] Charge conjugation invariance. 
\item[iii)] BRS-invariance:\\
 The BRS-transformations of the basic fields are defined as       
\begin{eqnarray} \label{y13}
A_{\mu}(x) &\longrightarrow &A_{\mu}(x) - \rho_{\mu}(x)\, \vep\, ,
    \nonumber \\
\psi ^j(x) &\longrightarrow & \,\psi ^j(x) - \rho ^j(x)\, \vep\, , \nonumber \\
\opsi^{\,j}(x) &\longrightarrow& 
\opsi^{\,j} (x) - \overline{\rho}^{\,j}(x)\, \vep\, ,   \\
c(x)& \longrightarrow& c(x)\,,\nonumber \\[-.2cm]
\bar{c}(x)& \longrightarrow& \bar{c}(x) -
 \frac{1}{\alpha}\, \partial_{\nu}A_{\nu}(x) \, \vep\ , \nonumber
\end{eqnarray}
\end{enumerate}  
using the composite fields (\ref{BRS}); $\,\vep\,$ is a 
Grassmann element not depending on space-time 
that commutes with the fields $A_{\mu}$
but anticommutes with 
the (anti-)fermions $\psi, \opsi$ and the (anti-)ghosts $c, \bar{c}\,$.\\
To show the BRS-invariance of the total classical action
(\ref{y9a}) one first observes that the composite classical fields
(\ref{BRS}) are themselves invariant under the BRS-transformations 
(\ref{y13}). Herewith it follows easily that 
the sum $ S_{\rm g.f.} + S_{\rm gh} $ is invariant under the 
transformation (\ref{y13}). Finally, on $ \,S_{\rm inv} $ act only 
the BRS-transformations of the fields $ A_{\mu} , \psi , \opsi \,$,
 which amounts to local gauge transformations. \\
We observe that upon scaling the composite fields (\ref{y13}) 
entering the BRS-transformations as well as $S_{gh}$ (\ref{y8}),
 by a factor of $ \la $, the corresponding $ S_{\rm BRS} $ remains 
invariant under such BRS-transformations.
BRS-invariance is considered to be
 sufficient for the gauge invariance of the S-matrix
if it exists \cite{Zi}. 
Note that contrary to electrodynamics, charge conjugation invariance 
does not forbid terms which are odd monomials in the gauge 
field.
The absence of such terms in QED is often termed Furry's theorem.\\ 
The fields $A_\mu,\overline{c},c$ and $\overline{\psi},\psi$ have 
 mass dimensions $1$ and $3/2$ respectively . We associate the ghost 
number $1$ to $\,c\,$, ghost number $-1$  to $\,\overline{c}\,$, 
and ghost number $0$ to $\,A_\mu,\,\overline{\psi},\,\psi\,$.
With these assignments the action has mass dimension 0 and ghost number
$0\,$.



\section{Regularized path integrals and renormalization \label{rere}}

In this section we shall introduce the path integral formulation of
chiral U(1) gauge theory. From momentum space regularized path 
integrals one derives the flow equations of the renormalization 
group on which renormalization theory in full generality can be based.
In fact the flow equations allow to deduce inductive
bounds on the Schwinger functions which imply renormalizability,
as was realized by Polchinski \cite{Po}, see also \cite{KKS}.
We try to be short on renormalization theory here since it is our aim
to confirm that the theory we consider {\it cannot be renormalized 
maintaining local gauge symmetry}. Using  flow equations it is
straightforward to see that it  {\it can be renormalized abandoning
local gauge symmetry}, a fact which one might state as
{\it renormalizability in the weak sense} \cite{Le}. Thus we will
not present the flow equations here, but just introduce the
 formalism on which they are based and from which we can
deduce the Slavnov-Taylor-identities (STI), which are violated in the presence
of cutoffs. We will then present the statements on renormalization 
theory we need in order to be able to verify whether the STI can
be restored on taking away the regulators. A complete presentation
of renormalization theory, in a case where the answer to this question
is affirmative, was presented in \cite{KM}.


\subsection{The regularized effective action \label{rea}}
 
\noindent
Bosonic field variables are generically denoted by 
$\phi$. Generally one may 
consider that  they are smooth 
functions\footnote{the support of Gaussian measures
depends on their regularity properties. For efficient regulators
our assumption turns out to be almost surely realized. 
For weaker forms of
regulators the subsequent expressions are still well-defined in the
support of the measure even if it exceeds the space of 
$C^\infty$-functions. We do not make explicit a finite-volume cutoff
since our statements on vertex and Schwinger functions hold in the
infinite volume limit. It is straightforward to  start by considering 
the  theory an a torus, see \cite{M}.}.
We will use the following concise notations~:
\begin{eqnarray*}
<\phi,\, \phi'>\, \equiv \int\! dx\,  \phi(x)\, \phi'(x)\, ,\ \
(\phi\ast \phi')(y)\, \equiv \int\! dx\,  \phi(x)\, \phi'(x-y) \ .
\end{eqnarray*}
As regards Fourier transforms we set 
\begin{eqnarray*}
&&\phi(x)=\int_p e^{ipx}\ \hat{\phi}(p)\quad
\mbox{ with}\quad \int_p \equiv \int \frac{d^4p}{(2\pi)^4}
\quad \Longleftrightarrow\quad
  \hat{\phi}(p)=\int dx\  e^{-ipx}\ \phi(x)\\
&&\Longrightarrow\quad 
 \frac{\delta}{\delta \phi(x)}=(2\pi)^4\int_pe^{-ipx}\ 
\frac{\delta}{\delta \hat{\phi}(p)}\ .
\end{eqnarray*}

 Quantization of the theory by means of functional integration 
  in the realm of (formal) power series 
 is based on a Gaussian measure
 related to the quadratic part (\ref{y10}) of $ S_{\rm BRS} $ (\ref{y9b}). 
 Denoting the differential operators  appearing with the various fields
as
 \begin{equation} \label{f1}
D_{\mu \nu} := (- \Delta +M^2 ) \,\delta_{\mu  \nu} -
 \frac{1-\alpha }{\alpha } \,\partial _{\mu}
  \partial _{\nu}\ ,\quad
i\,\slashed{\pa}_{ij}:= i\,\pa_{\mu} (\ga_{\mu})_{ij}\ ,\quad
D:=  - \Delta + \al M^2\ , 
\end{equation} 
 we write
\begin{equation} \label{f2}
 \int dx\  {\mathcal L}_{\rm quad} (x) =
 \frac12 \,\langle A^{a}_\mu, D_{\mu  \nu }\, A^{a}_\nu \rangle 
 + \,\langle \bar{\psi}, i \slashed{\partial}\,\psi  \rangle
  - \langle \bar{c},\, D\, c  \rangle \ .
  \end{equation}
  To the differential operators (\ref{f1}) are
 associated the (free) propagators
  \eq \label{f3}
  C_{\mu \nu} (x, y) = \int_k 
   e^{ik(x-y)}\,  C_{\mu \nu} (k) \, ,\ \,
S_{ij}(x,y) = \int_k e^{ik(x-y)}\,  S_{ij} (k) \, ,\ \,
  C (x, y) =\int_k    e^{ik(x-y)}\,  C (k) 
\eqe
with
    \begin{equation} \label{f4}
   C_{\mu  \nu}(k) = \frac{1}{k^2+M^2}
         \Big( \delta _{\mu \nu} - (1-\alpha )
\frac{k_\mu k_\nu}{k^2+\alpha M^2} \Big)\ , \
    S_{ij}(k) = \frac{\slashed{k}_{ij}}{k^2}\ , \ 
  C(k) = \frac{1}{k^2 +\al M^2}\ .
  \end{equation}
The Gaussian product measure is then defined with the aid of 
 covariances  which are a regularized version
 of the propagators (\ref{f3}), (\ref{f4}).
 We choose a cutoff function  
$\sigma_{\Lambda}(k^2) \,$
and set 
\begin{equation} \label{f5-1}
\sigma_{\Lambda,\Lao}(k^2) \,\equiv\,
\sigma_{\Lao}(k^2) \,-\, \sigma_{\La}(k^2)\ .
\end{equation}
For the bosons we may for example choose as in \cite{KM}
\begin{equation} \label{f5}
\sigma ^B_{\Lambda}(k^2) \,=\,
 \exp \Big ( -  \frac{(k^2 + M^2)(k^2 + \alpha M^2)( k^2)^2 }
{\Lambda^{8}} \,\Big )\ . 
\end{equation}
This cutoff function
 is positive, invertible and analytic, and has the property 
 \begin{equation} \label{f6}
\frac{d}{d k^2} \,\sigma ^B_{\Lambda}(k^2) |_{k^2 =0} \, = \, 0\ ,
 \end{equation} 
which is helpful in the analysis of  the STI in \cite{KM}.
For the fermions we choose a weaker cutoff more adapted 
for explicit 1-loop calculations of section \ref{PVCalc}. 
In fact we simply choose
\eq
 \sigma ^F_{\Lambda}(k^2) \,=\,
 \Big (1 -  \frac{k^2}{k^2 + \Lambda^{2}} \,\Big ) 
\end{equation}
 or higher powers thereof, i.e. a Pauli-Villars type cutoff.
 
Employing these cutoff functions we define the regularized propagators,
 with  UV-cutoff $ \Lao < \infty \,$ and a flow parameter $\La$ 
satisfying
   $ 0 \leq \Lambda \leq \Lambda_0  $,
\[
 C^{\Lambda,\Lambda_0}_{\mu \nu}(k) \equiv  
   C_{\mu \nu}(k)\  \sigma ^B_{\Lambda,\,\Lambda_0}(k^2)\ ,\quad
 C^{\Lambda,\Lambda_0}(k) \equiv  
   C(k)\  \sigma ^B_{\Lambda,\,\Lambda_0}(k^2)\ ,  
\]
 \begin{equation} \label{f7}
S^{\Lambda,\Lambda_0}(k) \equiv  
   S(k)\  \sigma ^F_{\Lambda,\,\Lambda_0}(k^2)\ . 
\end{equation}
It is convenient to introduce   a short collective notation
for the various fields and their sources:\\
i) We denote the physical fields and the corresponding
 sources, respectively, by
\begin{equation} \label{f9}
\varphi  = ( A_{\mu }\, , \, \psi ^j,\, \opsi^{j}) \
 ,\quad J = ( j_{\mu }\, , \,\ochi^{j}\, ,\,  \chi ^j)\ ,
\end{equation}
ii) and all fields and their respective sources by
 \begin{equation} \label{f10}
\Phi  = ( \varphi\, , \, c, \, {\bar c})  \ , 
               \quad  K = ( J\, ,\, {\bar \eta},\, \eta )\ .
 \end{equation} 
 The sources  $\chi ^j\,,\ \ochi^{j}\,$ and
$\, \eta\,, \ {\bar \eta}\,$ are Grassmann elements,
$\, \eta\,, \ {\bar \eta}\,$ have ghost
 number $+1$ and $-1$, respectively. In the sequel, we exclusively use
  left derivatives with respect to these quantities.\\
  The characteristic functional of the Gaussian product measure
 with the covariances
 from (\ref{f7}), (\ref{f4}) - multiplied by $\hbar$ in view of the 
loop expansion -   is  then given by
 \begin{equation} \label{f11}
\int d\mu_{\LLz} (\Phi ) \, e^{\, \frac{1}{\hbar} \langle \Phi , K \rangle}
         \,=\  e^{\, \frac{1}{\hbar} P^{\Lambda, \Lambda_0}(K)}\ ,
 \end{equation}
 where
 \begin{equation} \label{f12}
\vp(x) J(x) \equiv  A_{\mu}(x) j_{\mu}(x) \,+\, \ochi^{j}(x)\,\psi^j(x)
\,+\,  \opsi^{j}(x) \ \chi ^j(x)\ ,
\end{equation}
and 
 \begin{equation} \label{f12a}
\langle \vp  , J \rangle  \equiv \int\! dx\ \vp(x) J(x)
\ ,\quad 
\langle \Phi , K \rangle  \equiv \langle \vp  , J \rangle  +
   \int dx \Big(  \oc(x) \eta(x) + \oeta(x) c(x) \Big)\ ,
 \end{equation}
 \eq \label{f13}
 P^{\La,\Lao}(K)  \,=\, 
    \frac12 \,\langle j_{\mu}, C^{\LLz}_{\mu \nu} \,j_{\nu} \rangle 
     + \langle \ochi , S^{\LLz} \,\chi \rangle 
           - \langle \etb, C^{\LLz}\,\et \rangle \ .
 \eqe 

We now  consider the
 generating functional $ L^{\LLz} (\Phi ) $ of the
regularized  (through $\si_{\La,\Lao}\,$) connected amputated Schwinger
 functions (CAS) given by
  \begin{equation}
     \label{f14} 
     e^{-\frac{1}{\hbar}\left( L^{\Lambda ,\Lambda _0}(\Phi) 
               + I^{\Lambda ,\Lambda _0} \right)}
   =  \int d\mu _{\Lambda ,\Lambda _0}(\Phi ' )\  e^{-\frac{1}{\hbar}
                L^{\Lambda_0 ,\Lambda _0}( \Phi ' +\Phi ) }\ .
\end{equation}
We impose $L^{\Lambda , \Lambda _0}(0)  =  0\,$ so that 
the constant $ I^{\Lambda ,\Lambda _0} $ is the vacuum part
 which is proportional to the volume
because of translation invariance. 
It therefore requires  to consider the theory at first in a finite 
   volume $ \Omega \subset \mathbf{R}^4 $. For details see \cite{M}.

 Since the regularization necessarily violates the local gauge symmetry,
 the bare functional  $L^0(\Phi)=L^{\LLzz} (\Phi )$ 
in a first stage has to be chosen sufficiently general in order to allow
for  a finite limit $\Lao \to \infty\,$  at the end. We set
 \begin{equation} \label{f16}
   L^0(\Phi)=L^{\LLzz} (\Phi ) 
= \int dx \  \mathcal{L}_{\rm int} (x)\ +\
          L^{\LLzz}_{c.t.} (\Phi ) 
 \end{equation}
thus adding to the interaction part (\ref{y11})
of classical origin,  counter terms $ L^{\LLzz}_{c.t.} $, which a
priori include all local terms of mass dimension $\leq 4\, $
 permitted by the unbroken
{\it global} symmetries, i.e. Euclidean $O(4)$-invariance, charge
conjugation 
and global gauge invariance (\ref{y4}).
     There are six such 
terms, by definition all at least 
of order $\mathcal{O}(\hbar) $. The general bare
 functional can be written as follows~:
 \begin{eqnarray}
L_{c.t.}^{\Lao,\Lao}
&=&\int d^4x \Bigl[ 
(\Sigma^{\overline{\psi}\psi})^0 \,i\,\overline{\psi}\slashed{\partial}\psi
+\frac{(\delta M^2)^0}{2}\,A^2
+\frac{\Sigma_{long}^0}{2\alpha}\,(\partial A)^2
+\frac{\Sigma_{trans}^0}{4}\,F^2
\nonumber
\\
&&+\ \frac{(F^{AAAA})^0}{4!}\,(A^2)^2
+(\delta g)^0\,\overline{\psi}\slashed{A}\gamma_5\psi
   \Bigr]\ .
\label{L0}
\end{eqnarray}
In the abelian theory the ghosts 
are not coupled to the other fields.
It is therefore not necessary to introduce counter terms 
for the ghost fields. Note that a fermion mass term is 
not compatible with global gauge symmetry.\\
We also note that for $\La =\Lao$ (i.e. when the regularized
propagator vanishes), 
we have the intuitively obvious
equality between the generating functionals of the connected
and one-particle irreducible functions \cite{KM} denoted by $\Ga\,$
 \begin{equation} 
L_{c.t.}^{\Lao,\Lao}\,=\,\Gamma_{c.t.}^{\Lao,\Lao}\ .
 \end{equation}

\subsection{Inserted Schwinger functions \label{FEIS}}

To analyse the
 Slavnov-Taylor identities (STI), we have to consider 
Schwinger functions with a composite field inserted, too.
 Two kinds of such insertions have to be dealt with: local insertions
 implementing the BRS-variations, and a space-time integrated 
insertion representing the  violation of the STI.\\
The classical composite BRS-fields (\ref{BRS})  have mass 
dimensions $2$ and  $5/2$ (the latter if a fermion field appears). 
They transform
as axial vector, spinor and anti-spinor respectively, and they  
 have fermion number $0,\, \pm 1$ and ghost number $1$. Hence, 
allowing for counterterms, we introduce the bare composite 
fields\footnote{one may ask whether one should also introduce
a factor of $R_4^{0}$ 
for the BRS-transform of the antighost,
cf. the last relation in (\ref{BRS})~; such a factor is redundant
however because we may always choose an overall normalization
freely.}  
\alpheqn
\begin{eqnarray}
\label{3.12a}
& & \rho_{\mu}^{0}(x) 
= R_1^{0}
\, \partial_{\mu}\, c(x)\ ,\\
\label{3.12b}
& & \rho^{j,0}(x) 
= R_2^{0} \,ig\, (\ga_5 \psi(x))^j\, c(x) \ ,  
\\
\label{3.12c}
& & \bar{\rho}^{j,0}(x)  
= R_3^{0} \,ig\,(\opsi(x)\ga_5)^j \, c(x)\ ,
\end{eqnarray}
\reseteqn
keeping the  notation from (\ref{BRS}) but using
 it henceforth exclusively according to (\ref{3.12a})-(\ref{3.12c}). We set
 \begin{equation} \label{i2}
  R^{0}_{i} \, = \, 1+ \mathcal{O}(\hbar)\ , 
 \end{equation}
 thus viewing the counterterms again as formal power series in $\hbar $ ;
 the tree order ${\hbar}^0 $ provides the classical terms (\ref{BRS}). 
We note that the modified composite 
fields (\ref{3.12a})-(\ref{3.12c}) remain
 {\it invariant} under the BRS-transformations (\ref{y13}) 
if we employ the generalized composite fields
  (\ref{3.12a})-(\ref{3.12c})  in place of the original ones, 
(\ref{BRS}). Contrarily to the nonabelian case, 
this invariance does not enforce additional constraints on the 
$R_i^{0}$. 

  To generate Schwinger functions  with such insertions, 
the bare interaction (\ref{f16}) is modified adding 
 the composite fields (\ref{3.12a})-(\ref{3.12c}) coupled
 to corresponding sources
 \begin{equation} \label{i4}
\tilde{L}_0
=\tilde {L}^{\LLzz}(\rho ; \Phi) 
\equiv L^{\LLzz}(\Phi) + L^{\LLzz} (\rho )\ ,
 \end{equation}
  \begin{equation} \label{i5}
 L^{\LLzz} (\rho ) 
= \int dx\ \{ \zeta_{\mu}(x) \rho^{0}_{\mu}(x) +
        \bar{\zeta}^j(x) \rho^{j,0}(x)
 + \bar{\rho}^{j,0}(x) \zeta^j (x)\}\ . 
 \end{equation}
According to the properties of these composite fields, the
sources $\zeta_{\mu},\, \bar{\zeta}^j,\, \zeta^j\,$ are Grassmann elements,
they  have canonical dimension $2$ respectively $3/2$ for the last
two,  and ghost number 
$-1\,$.
For the insertions and their respective sources we also introduce a
short collective notation
\begin{equation} \label{i6}
   \rho = ( \rho_{\mu},\ \rho^j,\ \bar{\rho}^j )
 \ , \quad  \zeta = ( \zeta_{\mu}, \ \bar{\zeta}^j,\ \zeta ^{j}) \ .
 \end{equation}
 Using now (\ref{i4}) in place of $ L^{\LLzz} $ as the bare action in 
 the representation (\ref{f14}),  provides the functional 
 $ {\tilde L}^{\LLz}(\rho\,; \Phi)\, $, from which
 the generating functional of the regularized CAS
 with one  insertion $\rho^0_{\mu}(x) $ follows as
 \begin{equation} \label{i7}
 L^{\LLz}_{\zeta_\mu }(x \, ; \Phi ) 
\equiv \frac{\delta }{\delta \zeta_\mu (x)} 
          {\tilde L}^{\LLz}(\rho\,; \Phi)|_{\,\rho =0} \ , 
 \end{equation}
 and similarly for the other insertions from (\ref{i5}).
 In the infinite volume limit, and performing a Fourier
 transform of the insertion position we obtain
 \eq \label{i8}
 {\hat L}^{\LLz}_{\zeta_{\mu} }( q \, ; \Phi ) =
 \int dx \ e^{iqx}\ L^{\LLz}_{\zeta_{\mu} }(x \, ; \Phi ) \ .
 \eqe

 We shall describe  in Section 4, how the initial regularization,
 necessarily violating the STI, 
leads to another insertion which we denote as
\eq \label{i10}
 \tilde{L}^{\LLzz}(\theta )  \equiv  \int dx\,\theta (x)\, N(x)\, ,  \qquad
 N(x)  =  Q(x) + Q '( x\, ; \Lambda_0^{-1} )\ . 
\eqe
Here $\theta\,$ is another  source function.  
The individual terms of $ N(x)\, $ involve at most five fields 
and have ghost number $1$. Furthermore, $Q(x)$ is a local 
polynomial in the fields and their derivatives,
 having  canonical mass dimension $ D=5$, whereas
 $ Q '( x \,; \Lambda_0^{-1}) $ is nonpolynomial in the 
field momenta but suppressed by powers of ${\Lambda_0}^{-1}\,$.
In fact we will only need the spacetime integrated
insertion which is obtained form the local one via
functional derivation and subsequent integration.
We denote
\eq \label{i12}
   L^{\LLz}_{\theta} (\Phi )\,  
\equiv\,   \int dx\ L^{\LLz}_{\theta} (x\,; \Phi )
\, \equiv\,
\int dx\ \frac{\delta }{\delta \theta (x)} 
  {\tilde L}^{\LLz} (\theta\,; \Phi )|_{\theta =0}\ .
  \eqe

\subsection{Proper Vertex Functions \label{PVFS}}
Our analysis of the STI will be based 
on a representation in terms 
of proper vertex functions (1PI), since the extraction of relevant
parts from the STI is simpler and more transparent in terms of those
than in terms of the CAS. 
We will basically skip here the passage to the 1PI-functionals
which is performed explicitly in \cite{KM}, \cite{M}, and only 
give some basic results.

The field variables of the Legendre transformed functional 
are denoted through underlined variables
$\underline{A_\mu}, \underline{\psi}^j,  \underline{\opsi}^j,
 \underline{c}, \underline{\oc}\,$, and analogously for the collective
notations $  \underline{\vp},\  \underline{\Phi}$.
We can then obtain the generating functional of regularized
vertex functions
\[
\Ga ^{\La,\Lao}( \uFi)\ ,
\]
and also the corresponding generating functional of inserted 
regularized vertex functions
\[
{\tilde \Ga} ^{\La,\Lao}( \rho; \uFi)\ .
\]
 Since we restrict to perturbation theory, the generating functional
 will be considered within a formal loop expansion
  \begin{equation} \label{f19}
\Ga ^{\La,\Lao}( \uFi) =  \sum_{l=0}^{\infty} \hbar^l\
\Ga ^{\La,\Lao}_{l}( \uFi)\ .
 \end{equation}  
Furthermore, decomposing into particular $n$-point vertex functions
we introduce a multiindex $n$, the components of which  denote the
 number of each source field species appearing,
together with its modulus and its norm defined as follows~:
 \begin{equation} \label{f20}
  n = ( n_A,\,  n_\psi,\, n_{\opsi},\,n_c ,\, n_{\bar c} ) \, ,\
 |n| = n_A +n_\psi + n_{\opsi}\, + n_c + n_{\bar c} \, ,
||n|| = n_A +\frac32(n_\psi + n_{\opsi})  + n_c + n_{\bar c} \ .
 \end{equation}
The corresponding regularized 
 vertex functions in momentum space are then obtained
 through functional derivation
\eq
(2 \pi)^{4(|n|-1)} \de^{n}_{\uFi(p)}\Ga^{\La,\Lao}_{l}(\uFi)
  |_{\uFi \equiv 0}\,=\,
\de(p_1+\dots+p_{|n|})\, \Ga^{\La,\Lao}_{l,n}(p_1,\cdots,p_{|n|})\ ,
\label{cag1}
\eqe
\eq
(2 \pi)^{4(|n|-1)} \de^{n}_{\uFi(p)}\Ga^{\La,\Lao}_{\zeta_\mu\, ;\,l}(q;\,\uFi)
   |_{\uFi \equiv 0}\,=\,
\de(q+p_1+\dots+p_{|n|})\,\Ga^{\La,\Lao}_{\zeta_\mu ;\,l,n}
         (q ; p_1,\cdots,p_{|n|})\ ,
\label{cag2}
\eqe
\eq
(2 \pi)^{4(|n|-1)} \de^{n}_{\uFi(p)}\Ga^{\La,\Lao}_{\theta;\,l}(\uFi)
   |_{\uFi\equiv 0}\,=\,
\de(p_1+\dots+p_{|n|})\, \Ga^{\La,\Lao}_{\theta;\,l,n}(p_1,\cdots,p_{|n|})\ .
\label{cag3}
\eqe
 For the sake of a slim appearance, the notation does not reveal how
 the momenta are assigned to the multiindex $n$, 
 and in addition, the $ O(4) $-tensor structure remains
 hidden. By definition the $n$-point function 
 is completely symmetric (antisymmetric) if the variables
 that belong to each of the commuting (anti-commuting) species occurring are
  permuted.

\subsection{Weak renormalizability \label{WR}}

In this section we report on a number of results obtained
from renormalization theory based on flow equations, which 
we will need subsequently in the analysis of the 
STI. We try to be short in this respect since
it will turn out (as expected) that the model considered cannot be
renormalized as a gauge theory. 

With the aid of the flow equations one can deduce inductive
bounds on the Schwinger functions which imply renormalizability,
as was realized by Polchinski \cite{Po}, 
see also \cite{KKS}, and  \cite{KK} where the  flow equations for 
composite operators were introduced.
For a more recent presentation see \cite{M}. 
The facts necessary  to treat theories with massless fields can be
inferred from \cite{GK}, see also \cite{KK94}, \cite{KKQ},  and \cite{Le}.

As usual the relevant parameters of the theory have to be fixed
through renormalization conditions. The relevant part of the
functional $\Ga ^{0,\Lao}\,$ is analysed in \ref{relga}. In a 
(partially) massless theory marginal terms which are 
(logarithmically) infrared divergent by power-counting at zero 
momentum, have to be renormalized
at non-exceptional\footnote{i.e. no nontrivial subsum vanishes} external 
momenta. We therefore impose the following
renormalization conditions at any loop order $l\in \mathbb{N}$

\begin{eqnarray}
(\Gamma^{0,\Lambda_0}_{l,(0,2)})_{ij}(0)&=&0\ ,\qquad\qquad\qquad
\qquad\qquad\qquad\qquad\qquad\qquad\qquad\qquad\qquad\qquad  \\
\label{sips}
\partial_\mu
(\Gamma^{0,\Lambda_0}_{l,(0,2)})_{ij}(0)&=& 
\Sigma^{\overline{\psi}\psi}_l\ (\gamma_\mu)_{ij} \ , \\
(\Gamma^{0,\Lambda_0}_{l,(2,0)})
_{\mu\nu}(p_R) &=&
(\delta M^2)_l\, \delta_{\mu\nu} \,+\,
(p_R^2\,\delta_{\mu\nu} - p_{R,\mu} p_{R,\nu})\Sigma_{trans,l}\,+\,
\frac{p_{R,\mu} p_{R,\nu}}{\alpha}\,\Sigma_{long,l}\ ,   \\
(\Gamma^{0,\Lambda_0}_{l,(3,0)})_{\mu\nu\rho}(0)&=& 0 \ , \\
(\Gamma^{0,\Lambda_0}_{l,(4,0)})_{\mu\nu\rho\sigma}
({\vec p}^{\,\,(4)}_{R})
&=& \frac{F^{AAAA}_l}{3}\ ( \delta_{\mu\nu}\delta_{\rho\sigma}
+\delta_{\mu\rho}\delta_{\nu\sigma}+\delta_{\mu\sigma}\delta_{\nu\rho}) \ , \\
(\Gamma^{0,\Lambda_0}_{l,(1,2)})_{\mu ij}(0)
&=&(\delta g)_l\ (\gamma_\mu\gamma_5)_{ij} \ . 
\end{eqnarray}
 In (\ref{sips}) we derive with respect 
to the momentum associated to the field $\overline{\psi}$. 
We denote by $p_R=p_{1R} $ a fixed nonvanishing momentum.
Then the four momenta in 
${\vec p}_{R}^{\,\,(4)}= (p^{(4)}_{1R},\ p^{(4)}_{2R},
\ p^{(4)}_{3R},\ p^{(4)}_{4R}) $ 
  may be chosen  such that they point from the centre into the corners 
of a  tetrahedron - or similarly  into the corners of an equilateral triangle
in the case of the momenta ${\vec p}_{R}^{\,\,(3)}\,$ of
a three point function. 
From power counting one may also expect that 
$\partial_\sigma
(\Gamma^{0,\Lambda_0}_{l,(3,0)})_{\mu\nu\rho}(p_{1},p_{2},p_{3})$ 
contains  a relevant contribution, which then should
be proportional to the tensor $\vep_{\mu\nu\rho\sigma}\,$.
In fact the analysis of this term in section \ref{triang}, see in particular
(\ref{nor}), (\ref{expr}) excludes such a contribution.
Still this term is directly related to the anomaly in the STI,
see below (\ref{Z3}, \ref{r5}), (\ref{ano0}, \ref{ano}) 
and section \ref{triang}.
For inserted vertex functions, with $D$ being the dimension of the 
insertion, similarly all local terms of dimension $\le D\,$ have to be
fixed by renormalization conditions, where analogous restrictions on the
external momenta have to be observed. For the inserted functional
$\Ga_1$ appearing in the VSTI we have
$D=5$, and the corresponding relevant terms are listed explicitly 
in section \ref{relga1}.\\
With these renormalization conditions
the subsequent proposition holds for non-inserted vertex functions, 
if we start from the (inter)action (\ref{f16}),
where the counter terms are calculated as functions of the 
renormalization conditions. For inserted vertex functions 
it holds  with the same  conditions imposed on the noninserted
theory, and for a bare inserted functional calculated as before from
analogous renormalization conditions on the relevant inserted terms.
We state the proposition without proof, since its proof can be
inferred from \cite{KM}, \cite{GK}, \cite{Le}; knowing that the anomaly
will prevent us from making the corresponding statement 
on strong renormalizability (i.e. including the restoration of gauge
symmetry) anyway.  

\begin{prop}
\label{propv}
\textbf{Weak renormalizability of chiral U(1) gauge theory}\\
For fixed non-exceptional external momentum configurations
 $\vec p\,$ the vertex functions  
\eq \label{wr1}
{\Ga}^{\Lambda,\Lao}_{l ,\, n}( \vec{p}\,)
 \eqe
are uniformly bounded in $\Lao$. 
Furthermore the limits 
\[
\lim_{\Lao \to \infty} {\Ga}^{\Lambda,\Lao}_{l ,\, n}(\vec{p}\,)
\equiv  {\Ga}^{\Lambda}_{l ,\, n}(\vec{p}\,)
\]
and 
\[
\lim_{\La \to 0} {\Ga}^{\Lambda}_{l ,\, n}(\vec{p}\,)
\equiv  {\Ga}_{l ,\, n}(\vec{p}\,)
\]
exist and are smooth functions in the open set of non-exceptional momenta.\\
The same statements  also hold for  inserted vertex functions 
\eq \label{wr2}
{\Ga}^{\Lambda,\Lao}_{\zeta_\mu;\, l ,\, n}(q; \vec{p}\,)
 \eqe
and 
\eq \label{wr3}
{\Ga}^{\Lambda,\Lao}_{\theta;\, l ,\, n}(\vec{p}\,)\ .
 \eqe
\end{prop}

\noindent
It is also possible to control the singularities of the
vertex functions at exceptional momenta, see \cite{GK}.\\[.2cm] 
For the analysis of the possible restitution of the STI
in the renormalized theory the following 
 statement  on the
inserted functions  $\,{\Ga}^{\Lambda,\Lao}_{1;\, l ,\,
  n}(\vec{p}\,)\,$ is important (see \cite{KM})
\begin{prop}
\label{rest} 
\noindent
\textbf{Restitution theorem}\\  
If all renormalization constants imposed on the relevant
part of $\,{\Ga}^{0,\Lao}_{\theta;\, l }\,$ {\bf vanish}
and if possibly nonvanishing irrelevant  contributions to the bare 
functional $\,{\Ga}^{\Lao,\Lao}_{\theta;\, l }\,$ are 
bounded by $O\left(\Lao^{D-||n||-|w|}\,  
\mathcal{P}(\log(\frac{\Lambda_0}{\mu})\right)$ 
- for a suitable mass scale $\mu >0\,$- then  
for non-exceptional momenta the inserted functions 
\begin{eqnarray}
 \Gamma_{\theta;l,n}^{0,\Lambda_0}(\vec p)
\end{eqnarray}
 vanish in the limit 
$\Lao \to \infty$,  at least as $O\left(\Lao^{-1}\,  
\mathcal{P}\log\left(\frac{\Lambda_0}{\mu}\right)\right)\,$.\\  
The polynomials $\mathcal{P}$ 
have nonnegative coefficients which may depend on
$l,\,n,\,\mu,\, g,\, \al\,$,  
but not on $\, \Lambda\,,\, \Lambda_0\,$. 
\end{prop}

\noindent
Again we do not give a proof of this statement. In fact 
 {\it the presence of the anomaly turns out to be  an obstruction 
of its application on chiral U(1) gauge theory.}


\section{The Violated Slavnov-Taylor identities \label{viosti}}

\subsection{Deduction of the VSTI from the path integral \label{vstigen}}

 To examine the violation of the STI
produced by the UV cutoff $\Lambda_0$ we proceed in analogy with
\cite{KM}. We start from
 the generating functional of the regularized Schwinger functions
 at the value $\Lambda = 0$ of the flow
 parameter \footnote {again one should stay in finite volume
  as long as the vacuum part is involved}, 
\eq   \label{vs1} 
   Z^{0,\Lao}(K) \, = \, \int d\mu_{0,\Lao}(\Phi)
   \,e^{-\frac{1}{\hbar} L^{\LLzz}(\Phi) 
     +\frac{1}{\hbar} \langle \Phi , K \rangle }\  .
\eqe 
 The Gaussian measure $ d\mu _{0,\Lao}(\Phi ) $ corresponds
to the quadratic form $ \frac{1}{\hbar}\, Q^{0,\Lao}(\Phi ) $,
cf. (\ref{f11}), 
\begin{equation} \label{vs2}
  Q^{0,\Lao}(\Phi )  = 
   \frac12 \langle A_\mu, \big( C^{0,\Lao}\big)^{-1}_{\mu  \nu }
        A_\nu \rangle 
  +  \langle \opsi, (S^{0,\Lao})^{-1} \psi \rangle 
   -
  \langle {\bar c}, (C^{0,\Lao})^{-1} c \rangle .
\end{equation}  
 We define {\it regularized} BRS-variations
 (\ref{y13}), (\ref{3.12a})-(\ref{3.12c})
 of the fields by
\begin{eqnarray} \label{vs31}
\delta _{BRS} \,\vp(x) & = & - \,
 ( \sigma_{0,\Lao} \ast \rho)(x) \,\vep , \\  \label{vs32}
\delta _{BRS} \,\cb (x) & = & - \, \big (\sigma_{0,\Lao} \ast \frac{1}{\alpha }
 \,\partial _{\nu } \An \big ) (x) \,\vep \ .
 \end{eqnarray}  
The BRS-variation of the Gaussian measure has the form 
\begin{equation} \label{vs4}
 d\mu _{0,\Lao}(\Phi ) \mapsto  d\mu _{0,\Lao}(\Phi ) 
\Big( 1- \frac{1}{\hbar}\,
 \delta _{BRS}\, Q^{0,\Lao}(\Phi ) \Big)  \ .
\end{equation}
This BRS-variation of the Gaussian measure is obtained 
in the same way as for SU(2)-gauge theory, under the hypothesis
that there is no Jacobian stemming from the redefinition
of the field variables themselves. 
This is justified by
\begin{lemma}
\label{jac}
\noindent
We introduce a  cube of side length $L\,$ in $\mathbb{R}^4\,$
and expand the field variables in plane wave modes,
imposing periodic boundary conditions. We introduce
an UV cutoff $\Lao$ and restrict to wave numbers 
$k_n \in \ \mathbb{R}^4\,$ such that $|k_n| \le \Lao\,$.  
Imposing these regularizations 
the Jacobian associated with the change of variables
(\ref{vs31}, \ref{vs32}) equals 1. 

The elementary proof of this statement is in App. \ref{appa}.
From the proof it is  quite evident that the statement 
 holds for larger classes of regulators and mode 
expansions. This is in some sense opposed to the deduction
of the anomaly by Fujikawa \cite{Fu}, \cite{FuS} who relates 
it to a nontrivial Jacobian. On the other hand 
a statement analogous to ours can be found  \cite{GJ}, sect. II.A.  
We comment on Fujikawa's argument in App. \ref{fuji}. 
\end{lemma}

Inspecting (\ref{vs2}) we observe that the factor $\sigma_{0,\Lao} $ of
 the BRS-variations (\ref{vs31}, \ref{vs32}) just cancels its inverse entering
 the inverted propagators. 
Hence, the BRS-variation 
 of the Gaussian measure has mass dimension $ D = 5$. 
Invariance of the  regularized generating functional 
$Z^{0,\Lao}(K)$, (\ref{vs1}) 
 under the  BRS-variations  (\ref{vs31}, \ref{vs32})
then provides the {\it violated Slavnov-Taylor identities}
 \begin{equation} \label{vs5}
    0 \stackrel{!}{=}
 \int d\mu_{0,\Lao}(\Phi) \  e^{-\frac{1}{\hbar} L^{\LLzz}(\Phi)
 +\frac{1}{\hbar} \langle \Phi , K \rangle }
 \Bigl( \delta _{BRS}\, \langle \Phi , K \rangle - \delta _{BRS}\,
        ( Q^{0,\Lao} + L^{\LLzz}) \Bigr)\ .
 \end{equation}  
 The BRS-variations appearing in (\ref{vs5}) can be  dealt with,
 considering corresponding modified generating functionals,
where the notations are chosen as in \ref{FEIS}: \\
 i) The modified bare interaction (\ref{i4}) is defined
 \begin{equation} \label{vs6}
{\tilde Z}^{0,\Lao} (K,\rho ) \equiv \int d\mu_{0,\Lao}(\Phi)
    \,e^{-\frac{1}{\hbar} {\tilde L}^{\LLzz}(\rho ; \Phi) 
+\frac{1}{\hbar} \langle \Phi , K \rangle }\ .
 \end{equation} 
ii) The BRS-variations of the bare action and of the Gaussian measure 
 \begin{equation} \label{vs8}
 L^{\LLzz}_\theta \vep \equiv \,
      - \delta _{BRS} \Big( Q^{0,\Lao} + L^{\LLzz} \Bigr) 
         = \int dx\,  N(x) \,\vep 
 \end{equation} 
 form  a space-time integrated insertion with ghost number $1$.
 The variation of $ L^{\LLzz} $, however, keeps the regularizing 
 factor $ \sigma_{0,\Lao}$ of  (\ref{vs31}, \ref{vs32}), 
thus the integrand $ N(x) $ 
is no longer  a polynomial in the fields and their derivatives. 
 We  treat the integrand $ N(x) $ as a local insertion
  with a source $\theta(x)$, cf. (\ref{i12}). Introducing the corresponding
 bare action  $ {\tilde L}^{\LLzz}(\theta ; \Phi)\, $, we define  the
 functional \footnote{ Abusing notation we let the variables $\theta$ and
 $\rho$, respectively, denote different functions.}  
${\tilde Z}^{\Lz} (K,\theta ) $ in analogy to (\ref{vs6}). \\ 
 In terms of these modified $Z$-functionals
 the VSTI
 (\ref{vs5}) can now be written 
 \begin{equation} \label{vs9}
  \mathcal{D}_{0,\Lao} \, \tilde{Z}^{0,\Lao} 
(K, \rho ) |_{\,\rho =0} \, = \, 
   \int dx\ \frac{\delta }{\delta \theta (x)}
 \tilde{Z}^{0,\Lao} (K, \theta ) |_{\,\theta =0}\ . 
 \end{equation} 
Here we introduced a {\it regularized}  
BRS-operator\footnote{$\langle J \, , \frac{\delta }{\delta {\zeta}}\rangle$
is short for $\int dx\, \{j_\mu(x) \,\de_{\zeta_\mu(x)}
+\ochi^j(x)  \, \de_{\bar{\zeta}^j(x) }
- \de_{\zeta^j}(x) \, \chi^j(x)  \} \,$.}
 \begin{equation} \label{vs7}
 {\mathcal D}_{0,\Lambda_0} = 
 \big \langle J \, , \sigma_{0,\Lao}\, \frac{\delta }{\delta {\zeta}}
             \big\rangle
 +\big\langle \, \frac{1}{\alpha }
 \,\partial _{\nu} \frac{\delta }{\delta \jn}
    \, ,\sigma_{0,\Lao} \et \big\rangle\  .
 \end{equation} The modified $Z$-functional (\ref{vs6}) 
is related to the corresponding
   generating functional of modified CAS by
   \footnote{ The vacuum part $ I^{\Lz} $ is the same as in the case
 without insertion, since the insertion has ghost number 1} 
 \eq \label{vs10}
 {\tilde Z}^{0,\Lao} (K,\rho )  
=  e^{\frac{1}{\hbar} P^{0,\Lao}(K) } \,
  e^{-\frac{1}{\hbar} ( {\tilde L}^{0,\Lao}
 ( \rho ; \,\vp , \, c, \,\bar c ) + I^{\Lz}) } \ ,
 \eqe
 and analogously in case of ${\tilde Z}^{\Lz} (K,\theta ) $.
  Starting from the relations
between the generating functionals $\ti Z$ and the corresponding
generating functionals of the vertex-functions we can convert  (\ref{vs10})
at the value $ \La = 0 $  into the 
\emph{violated Slavnov-Taylor identities for proper vertex functions},
on  substituting there the fields $ \Phi $ by the underlined fields
 $ \underline{\Phi} $ which are the variables of the Legendre
 transform.
We obtain
\begin{equation} \label{vs22}
\Gamma^{0,\Lao} _\theta ( \underline{\vp}\, , \uca, \ucb )
 \, =\,
\Big\langle \frac{\delta \Gamma^{0,\Lao} }{\delta \underline{\vp}} 
  \, ,\sigma_{0,\Lao} \Gamma^{0,\Lao} _{\zeta} \Big\rangle 
            - \Big\langle \frac{1}{\alpha }\, \partial _{\nu} \uAn\,,\,
 \sigma_{0,\Lao} \frac{\delta \Gamma^{0,\Lao} }{\delta \ucb}\Big\rangle 
              \end{equation} 
 with
 \begin{equation} \label{vs23}
   \Gamma^{0,\Lao} _\theta (\underline{\vp} , \uca, \ucb ) \, = \, 
    L^{0,\Lao}_1 ( \vp , \ca, \cb ) \ .
 \end{equation}

\noindent
We rewrite the VSTI (\ref{vs22}) more explicitly as  
\begin{eqnarray}
 \Gamma_{\theta}^{0,\Lambda_0} \,= \,
-\frac{1}{\alpha}\,\langle  
\frac{\delta \Gamma^{0,\Lambda_0}(\underline{\Phi})}
{\delta \overline{\underline{c}}},\,
\sigma_{0,\Lao}\ast \partial \underline{A}\rangle
\ +\
\langle \sigma_{0,\Lao}\ast 
\frac{\delta \Gamma^{0,\Lambda_0}(\underline{\Phi})}
{\delta \underline{A}_\nu}, \,
\Gamma^{0,\Lambda_0}_{\zeta_\nu}(\underline{\Phi})\rangle 
\qquad\qquad\qquad\quad
\label{sti}\\
 -\ \langle \sigma_{0,\Lao}\ast
\frac{\delta \Gamma^{0,\Lambda_0}(\underline{\Phi})}
{\delta \underline{\psi}},\,  
\Gamma^{0,\Lambda_0}_{\bar\zeta}(\underline{\Phi})\rangle
\ -\  \langle \Gamma^{0,\Lambda_0}_{\zeta}
(\underline{\Phi}),\, \sigma_{0,\Lao}\ast
\frac{\delta \Gamma^{0,\Lambda_0}(\underline{\Phi})}
{\delta \overline{\underline{\psi}}} \rangle
\ .
\nonumber
\end{eqnarray}
$\,\Gamma_{\theta}^{\Lambda,\Lambda_0}\,$ 
represents  the violation of the STI.  
Due to the fact that the BRS-transform increases the dimension
of a monomial in the fields by one unit,  $\Gamma_{\theta}\,$ has to be
interpreted as the generating functional of  1PI-functions 
carrying an operator insertion of dimension 5 and ghost number one.  
Therefore relevant terms in this functional have mass
dimension $\le 5\,$.
Still following \cite{KM} we now 
analyse the relevant contributions  to the VSTI (\ref{sti})
in \ref{relvsti}\footnote{In \cite{KM} we also analysed the 
VSTI at the bare side, i.e. at $\La =\Lao\,$, in order
to verify the corresponding boundary conditions for Proposition \ref{rest}.
Since here we will show that the anomaly prevents us from
verifying the required boundary conditions at $\La=0\,$,
this second step becomes obsolete.}.

\subsection{The relevant contributions to the VSTI \label{relvsti}}

\subsubsection{The relevant  contributions to the 
functional $\Ga$  \label{relga}}

The generating functional $\Gamma^{0,\Lambda_0}$ is invariant under the 
Euclidean group, under charge conjugation and under global (chiral)
gauge transformations. 
We start listing  the contributions to 
 the relevant part of the generating functional 
$\Gamma^{0,\Lambda_0}$ i.e. those terms of mass dimension $\leq 4$
which respect these symmetries. We do not underline field variables 
nor do we indicate the dependence on $\Lambda_0\,$ 
or the loop-order $l\,$.

\begin{itemize}
\item
\begin{eqnarray*}
\Gamma_2=\int_p A_\mu(p) A_\nu(-p) \Gamma^{AA}_{\mu\nu}(p)
+\overline{\psi}^i(p)\Gamma_{ij}^{\overline{\psi}\psi}(p)\psi^j(-p)
-\overline{c}(p)c(-p)\Gamma^{\overline{c}c}(p)
 \end{eqnarray*}

\begin{eqnarray*}
\Gamma^{AA}_{\mu\nu}(p)&=&\frac{1}{2}\left[(M^2+\delta
  M^2)\delta_{\mu\nu}
+(p^2\delta_{\mu\nu}-p_\mu p_\nu)(1+\Sigma_{trans})
+\frac{p_\mu p_\nu}{\alpha}(1+\Sigma_{long})\right]\\
\Gamma^{\overline{\psi}\psi}(p)
&=&-\slashed{p}(1+\Sigma^{\overline{\psi}\psi})\\
\Gamma^{\overline{c}c}(p)&=&p^2+\alpha M^2\ .
 \end{eqnarray*}
 Since ghosts do not interact with other fields nor with themselves, 
the renormalization procedure does not modify the expression of their 
propagator. We have
\[
\Sigma^{\overline{\psi}\psi},\ \Sigma_{trans},\
\Sigma_{long},\ \delta M^2\, = \,O(\hbar)\ .
\]
Due to global gauge invariance there is no mass term in 
$\Gamma^{\overline{\psi}\psi}\,$. 

\item 
\begin{equation*}
\Gamma_3=\int_{p,q}\overline{\psi}^{\,i}(p)
\Gamma_{\mu,ij}^{\overline{\psi}A\psi}
(p,q)\psi ^j(q)A_\mu(-p-q)+A_\mu(p)A_\nu(q)A_\rho(-p-q)\Gamma^{AAA}_{\mu\nu\rho}
(p,q)
 \end{equation*}

\begin{eqnarray*}
\Gamma_{\mu,ij}^{\overline{\psi}A\psi}(p,q)&=&(\gamma_\mu
\gamma_5)_{ij} 
F^{\overline{\psi}A\psi},\qquad  F^{\overline{\psi}A\psi}=g + \delta
g\ ,\\
 \end{eqnarray*}
 with $\,\delta g,\, \Gamma^{AAA}_{\mu\nu\rho} = O(\hbar)\,$. The structure of 
$\Gamma_{\mu\nu\rho}^{AAA}$ is analysed 
in  section \ref{triang}.

 \item 
 \begin{eqnarray*}
\Gamma_4=\int_{p,q,r} \frac{1}{4!} 
A_\mu(p) A_\nu(q) A_\rho(r) A_\sigma(-(p+q+r))
\Gamma^{AAAA}_{\mu\nu\rho\sigma}(p,q,r) \ ,
 \end{eqnarray*}

 \begin{eqnarray*}
\Gamma^{AAAA}_{\mu\nu\rho\sigma}(p,q,r)=\frac{F^{AAAA}}{3}(\delta_{\mu\nu}
\delta_{\rho\sigma}+\delta_{\mu\rho}\delta_{\nu\sigma}
+\delta_{\mu\sigma}\delta_{\nu\rho}) \ ,
 \end{eqnarray*}
  with $F^{AAAA}=  O(\hbar)$.

\end{itemize}

\subsubsection{The relevant  contributions to the functional $\Ga_\theta$  
\label{relga1}}

Expanding $\Gamma_\theta$ up to terms of mass dimension 5 in fields and momenta 
 we obtain the relevant terms which are listed below. We first write
the corresponding contribution to the (V)STI for the corresponding
field content and then the relation which follows if one imposes the
corresponding relevant part of $\,\Ga_{\theta}\,$ to vanish. This
relation is expressed in terms of the momenta and of the 
renormalization constants. The field content is indicated in 
the upper index of $\Ga_{\theta}\,$.
 

\begin{enumerate}

\item 
\begin{eqnarray}\label{Z1}
\Gamma_{\theta}^{A_{\mu}c}(p,-p)
=\left( \frac{1}{\alpha}    
ip_{\mu}
\Gamma^{\overline{c}c}(p)-2iR_1p_{\nu}\Gamma^{AA}_{\nu\mu}(p)
 \right)
\end{eqnarray}
\begin{eqnarray}
\label{r1}
0&=^{^{\!\!\!\!\! !}}&
iM^2p_{\mu}\left(1-R_1(1+\frac{\delta M^2}{M^2})  \right)\ ,\\
\label{r2}
0&=^{^{\!\!\!\!\! !}}& 
i p_R^2\, p_{R,\mu}\frac{1}{\alpha}\left(  1-R_1(1+\Sigma_{long}))
\right)\ . 
\end{eqnarray}

\item 
\begin{eqnarray}\label{Z2}
\Gamma_{\theta}^{\overline{\psi}_i\psi_jc}(p_1,p_2,-p_1-p_2)=
-iR_1p_{3\mu}\Gamma^{\overline{\psi}A\psi}_{\mu,ij} 
-igR_3\left(\gamma_5\Gamma^{\overline{\psi}\psi}(-p_2)\right)_{ij}\\  
\nonumber
+igR_2\left(\gamma_5\Gamma^{\overline{\psi}\psi}(p_1)\right)_{ij}
\end{eqnarray}
\begin{eqnarray}
\label{r3}
0&=^{^{\!\!\!\!\! !}}& 
i (\slashed{p}_1\gamma_5)_{ij}\left[R_1   \left(g+\delta g \right)  
-gR_2(1+\Sigma ^{\overline{\psi}\psi}) \right] \ , \\
\label{r4}
0&=^{^{\!\!\!\!\! !}}& i (\slashed{p}_2\gamma_5)_{ij}
\left[R_1   \left(g+\delta g\right)-gR_3(1+\Sigma
  ^{\overline{\psi}\psi})
  \right] \ .  
\end{eqnarray}

\item
\footnote{We use the notation
$w \equiv (w_{1,1} , \cdots , w_{n-1,4} )\, , \ w_{i,\mu } 
\in \mathbf{N}_0\,  , 
  \ \partial ^{\,w} \equiv \prod_{i,\mu} 
  \Bigl(\frac{\partial }{\partial p_{i,\mu }} \Bigr)^{w_{i,\mu }} \, , 
 \ |w| \equiv \sum_{i,\mu } w_{i,\mu }\, .
$}
 
\begin{eqnarray}\label{Z3}
\Gamma_{\theta}^{ A_{\mu_1}A_{\mu_2}c}(p_1,p_2,p_3)
=-3!\,g\, R_1\,p_{3\mu} \Gamma^{AAA}_{\mu\mu_1\mu_2}(p_3,p_1)\ ,\quad
p_3\equiv -p_1-p_2
\end{eqnarray}
\begin{eqnarray}\label{r5}
0\, =^{^{\!\!\!\! !}}\
\partial^w\left( p_{3\mu} 
\Gamma^{AAA}_{\mu\mu_1\mu_2}(p_3,p_1)\right)
\Bigr|_{p_i \equiv {p}_{R,i}^{(3)}}   
\ ,\quad |w| \le 2\  .
\end{eqnarray}

\item 
\begin{eqnarray}\label{Z4}
\Gamma_{\theta}^{\overline{\psi}_i\psi_jA_{\mu_1}c}(p_1,p_2,p_3,p_4)
=-ig\left(R_3-R_2\right)
\left(\gamma_5\Gamma_{\mu_1}^{\overline{\psi}A\psi}\right)_{ij}\
,\quad
p_4=-p_1-p_2-p_3
\end{eqnarray}
\begin{eqnarray}
\label{r6}
0=^{^{\!\!\!\!\! !}}g(g+\delta g)\left(R_3-R_2\right)\ .
\end{eqnarray}

\item 
\begin{eqnarray}\label{Z5}
\Gamma_{\theta}^{A_{\mu_1} A_{\mu_2} A_{\mu_3} c}(p_1,p_2,p_3,p_4)
=-4!\,R_1\,
p_{4,\mu}\, \Gamma^{AAAA}_{\mu\mu_1\mu_2\mu_3}(p_1,p_2,p_3,p_4)
\end{eqnarray}
\begin{eqnarray}
\label{r7}
0\,=^{^{\!\!\!\!\! !}}\ \partial^w \left(p_{4,\mu}\
\Gamma^{AAAA}_{\mu\mu_1\mu_2\mu_3}
(p_1,p_2,p_3,-p_1-p_2-p_3)\right)\Bigr|_{p_i \equiv {p}_{R,i}^{(4)}}
\ ,\quad |w| \le 1\  .
\end{eqnarray}
\end{enumerate}
The subsequent five relations on the renormalization conditions
allow to verify the conditions indicated behind them 
\begin{eqnarray}
&&R_2=R_3  \qquad\qquad\qquad\qquad\quad
\Rightarrow\quad  (\ref{r6})\ ,\label{bel1}\\
&&R_1=\frac{1}{1+\Sigma_{long}} \qquad\qquad\qquad \Rightarrow\quad
(\ref{r2})\ ,\label{bel2}
\\
&&\frac{\delta M^2}{M^2}=\Sigma_{long}
\quad\qquad\qquad\qquad \Rightarrow\quad (\ref{r1})\label{bel4}\ ,\\
&&R_1=(1+\Sigma^{\overline{\psi}\psi})\,\frac{g}{g+\delta
  g}\, R_2 \quad\ \, \Rightarrow\quad (\ref{r3})\, ,\ (\ref{r4}) \ , 
\label{bel3}\\
&&F^{AAAA}=0 \quad\qquad\qquad\qquad \ \ \, \Rightarrow\quad
(\ref{r7}) 
\label{bel5}\ .
\end{eqnarray}
The last relation is sufficient to ensure (\ref{Z5})
since the tensor structure of $\Gamma^{AAAA}_{\mu\mu_1\mu_2\mu_3}\,$
implies that the higher order contributions in an expansion around
${\vec p}_R^{\,\,(4)}\,$ are irrelevant
\footnote{
\begin{eqnarray*}
\Gamma^{AAAA}_{\mu\nu\rho\sigma}(p_1,\ldots,p_4)
&=&(\delta_{\mu\nu}\delta_{\rho\sigma}+\delta_{\mu\rho}\delta_{\nu\sigma}
+\delta_{\mu\sigma}\delta_{\nu\rho})F^{AAAA}
+p'^2
(\delta_{\mu\nu}\delta_{\rho\sigma}+\delta_{\mu\rho}\delta_{\nu\sigma}
+\delta_{\mu\sigma}\delta_{\nu\rho})F_1\\
&+&\left[ p'_{1\mu}p'_{2\nu}\delta_{\rho\sigma}
+p'_{1\mu}p'_{3\rho}\delta_{\nu\sigma}
+p'_{1\mu}p'_{4\sigma}\delta_{\nu\rho}
+p'_{2\nu}p'_{3\rho}\delta_{\mu\sigma}
+p'_{2\nu}p'_{4\sigma}\delta_{\mu\rho}
+p'_{3\rho}p'_{4\sigma}\delta_{\mu\nu} \right]F_2\\
& +& O(p'^4)
\end{eqnarray*}
with $p'=p-p_R$.
}.
A simple solution of  (\ref{bel1}) to (\ref{bel5}) is given by
imposing the value 0 for all quantities of order $\hbar\,$. 
The tensor structure and the one-loop contributions
to $\Gamma^{AAA}_{\mu\nu\rho}$ are analysed in section \ref{triang}. 
Subsequently we  just write  $\,\Gamma_{\mu\nu\rho}\,$.
As a consequence of explicit calculation we find

\begin{prop}
\label{triangle} 
\noindent
 For $\Lambda=0$ and $\Lao < \infty\,$ the contracted 
three-photon-amplitude has the Feynman parameter representation
(denoting $d\mu_5 = \prod_{i=1}^5 dx_i \de(1-\sum_i x_i) \,$)
\begin{eqnarray}
\label{ano0}
p_{1\mu} \Gamma_{\mu\nu\rho}^{0,\Lambda_0}=\frac{2}{\pi^2} 
\epsilon_{\nu\rho \alpha\beta}  p_{2\alpha}p_{3\beta} \int 
 d\mu_5\,
\frac{x_3 \Lambda_0^6}{\left[\tilde{x}_{25}p_2^2
+\tilde{x}_3p_3^2+2x_{25}x_3p_2\cdot p_3+x_{123}\Lambda_0^2\right]^3}\ .
\end{eqnarray} 
The integrand and its up to second derivatives are absolutely integrable.
In the UV limit the integral  converges 
uniformly in momentum space and is given by 
\begin{eqnarray}
\label{ano}
\lim_{\Lambda_0\rightarrow \infty}
\left(p_{1\mu}\Gamma_{\mu\nu\rho}^{0,\Lambda_0}\right)
=\frac{2}{\pi^2}\,\epsilon_{\nu\rho\alpha\beta }\,p_{2\alpha}\,p_{3\beta}
\int  d\mu_5 \, \frac{x_3}{x_{123}^3} \ =\
\frac{1}{6\pi^2}\,\epsilon_{\nu\rho\alpha\beta }\,p_{2\alpha}\, p_{3\beta}\ .
\end{eqnarray}
\end{prop}

\noindent
Furthermore, we have the bounds
\begin{eqnarray*}
 | \partial^w\left( p_{1\mu}\Gamma_{\mu\nu\rho}^{0,\Lambda_0}
\,-\, p_{1\mu}\Gamma_{\mu\nu\rho}^{0,\infty}\right) |
\,=\, O\left(\frac{p^{4-|w|}}{\Lambda_0^2}\right)\ , \quad |w|\leq 2\ .
\end{eqnarray*}

\noindent
As a consequence the relevant part of the STI given through thr r.h.s.
of (\ref{Z3}) and (\ref{r5}) cannot be made vanish for any choice 
of renormalization point. In fact, as is explained in App. \ref{tensy},  
there is no relevant local term corresponding to the
three-photon-amplitude, and its second derivatives (at any non-exceptional
momenta) do not identically 
vanish according to (\ref{ano}). For  more details
on the three-photon-amplitude see App. \ref{triang}.  
Our conclusion is 

\begin{thm}\label{thm1}
The chiral U(1) gauge theory given through the Lagrangian
(\ref{y1}) is not renormalizable in the strong sense, that is 
to say such that the Slavnov-Taylor-Identities are restored in the 
renormalized theory. The obstruction is due to a nonvanishing
relevant (in the sense of the renormalization group) contribution
of the three-vector-boson amplitude $\Gamma^{AAA}_{\mu\nu\rho}$
 violating these identities so that the  STI violating functional
(\ref{vs23}, \ref{sti}) satisfies
\begin{eqnarray}
\left.[\Gamma^{0,\Lao}_\theta] \right|_{ins} \neq 0\ ,
\end{eqnarray}
for all (large) values of the UV cutoff $\Lambda_0\,$. 
\end{thm}
In fact the anomaly is closely related to 
the infrared singular behaviour of the (derivatives of) the
three-photon-amplitude. If the amplitude were analytic around 
zero momentum, the anomaly could not appear, as is explained 
in the next section and follows from  Lemma \ref{syle}.\\    

\noindent
{\bf Acknowledgement:} Ch. K.  would like to thank Stefan Hollands
for several discussions, in particular  on the heat kernel in
background fields.

\appendix
\section{Analysis of the three-photon amplitude \label{triang}}

\subsection{Tensor structure, symmetries\label{tensy}}

The three-(axial) photon-amplitude
\[
\Gamma_{\mu\nu\rho}(p_1,p_2,p_3)
\]
 is a tensor w.r.t. the euclidean $O(4)$-group. 
Euclidean symmetry, the parity
transformation of the axial vector field - 
which enforces the appearance of the Levi-Civita tensor 
$\,\epsilon_{\alpha\beta\gamma\delta}\,$ -
and translation invariance  which implies 
$p_1+p_2+p_3 = 0\, $,
permit to obtain the following decomposition
of the  tensor $\,\Gamma_{\mu\nu\rho}(p_1,p_2,p_3)\,$ 
in invariants~:
\begin{eqnarray*}\label{AAAA}
&&\Gamma_{\mu\nu\rho}(p_1,p_2,p_3)
=
A_1(123)p_{1\tau}\epsilon_{ \tau \mu\nu\rho}
+A_2(123)p_{2\tau} \epsilon_{ \tau \mu\nu\rho}\\
&&+A_3(123)p_{1\nu} p_{1\alpha} p_{2\beta} \epsilon_{ \alpha\beta \mu\rho}
+A_4(123)p_{2\nu} p_{1\alpha} p_{2\beta} \epsilon_{ \alpha \beta \mu\rho}
+A_5(123)p_{1\mu} p_{1\alpha} p_{2\beta} \epsilon_{\alpha \beta \nu\rho}\\
&&+A_6(123)p_{2\mu} p_{1\alpha} p_{2\beta} \epsilon_{ \alpha \beta \nu\rho}
+A_7(123)p_{1\rho} p_{1\alpha} p_{2\beta} \epsilon_{ \alpha \beta \mu\nu}
+A_8(123)p_{2\rho} p_{1\alpha} p_{2\beta} \epsilon_{ \alpha \beta \mu\nu }\ .
\end{eqnarray*}
Here we use the shorthand notation $A_i(\sigma(1)\sigma(2)\sigma(3))\,$ for
$\,A_i(p^2_{\sigma(1)},p^2_{\sigma(2)},p^2_{\sigma(3)})\,$,
and the $A_i\,$ are euclidean scalars.
Using also complete Bose symmetry  w.r.t. the 6 permutations 
of 
\[
(p_1,\mu)\,,\ (p_2,\nu)\,,\ (p_3,\rho)
\] 
and considering the  momentum configurations,
values of the tensor indices and identities between 
tensor components following from Bose symmetry as indicated in
(\ref{1}, {\ref{2}, \ref{3})
\begin{itemize}
\item 
\begin{eqnarray}
\label{1}
&&p_1=(p_{11},0,p_{13},p_{14}),\  p_2=(0,0,0,p_{24}),
\quad (\mu,\nu,\rho)=(1,1,2),(1,2,1),\\ \nonumber
&& \Gamma_{\mu\nu\rho}(p_1,p_2,p_3)=\Gamma_{\nu\mu\rho}(p_2,p_1,p_3)\
, \\
\nonumber
\end{eqnarray}
\item
\begin{eqnarray}
\label{2}
&&p_1=(p_{11},0,0,p_{14}),\ p_2=(p_{21},0,0,p_{24}),
\quad (\mu,\nu,\rho)=(1,2,3),\\
\nonumber&&\Gamma_{\mu\nu\rho}(p_1,p_2,p_3)
=\Gamma_{\nu\mu\rho}(p_2,p_1,p_3)=\Gamma_{\rho\nu\mu}(p_3,p_2,p_1)
=\Gamma_{\mu\rho\nu}(p_1,p_3,p_2)\ , 
\end{eqnarray}
\item
\begin{eqnarray}
\label{3}
&&p_1=(0,0,p_{13},p_{14}),\ p_2=(0,0,p_{23},p_{24}),
\quad (\mu,\nu,\rho)=(1,2,3),\\
\nonumber&&\Gamma_{\mu\nu\rho}(p_1,p_2,p_3)=\Gamma_{\nu\mu\rho}(p_2,p_1,p_3)\ ,
\end{eqnarray}
\end{itemize}
and solving them in terms of the eight invariants $A_i$, 
we obtain the following relations:
\begin{eqnarray}
\label{nor}
&&A_1(123)+A_1(231)+A_1(312) = 0  \\
&&A_1(123)=A_1(321\nonumber \\ 
&&A_2(123)=-A_1(213)\nonumber\\
&&A_5(123)=-A_4(213)\nonumber\\
&&A_6(123)=-A_3(213)\nonumber\\
&&A_6(123)=A_8(213)\nonumber\\
&&A_7(123)=A_4(231)\nonumber\\
&&A_3(123)=A_4(123)-A_4(321)\ .
\end{eqnarray}
Thus 
$\,\Gamma_{\mu\nu\rho}(p_1,p_2,p_3)\,$ can be expressed in terms
of only two amplitudes, for example $A_1=A,\ A_4=B\,$:
\begin{eqnarray}{\label{expr}}
\Ga_{\mu \nu\rho}(p_1,p_2,p_3)
&\,=\,&[ A(1,2,3)\, p_1^\tau -A(2,1,3) p_2^\tau \, ] \vep_{ \tau \mu\nu\rho} \\
&\,+\,& [B(1,2,3)p_{3\nu}+B(3,2,1)p_{1\nu}]\nonumber
\epsilon_{\al \be \rho\mu }p_1^\al p_2^\be  \nonumber\\
&\,+\,&  [B(3,1,2)p_{2\mu}+B(2,1,3)p_{3\mu}]
\epsilon_{\al \be \nu\rho}p_1^\al p_2^\be\nonumber\\  
 &\,+\,& [B(2,3,1)p_{1\rho}+B(1,3,2)p_{2\rho}]
\epsilon_{\al \be\mu\nu}p_1^\al p_2^\be   \ . 
\end{eqnarray}
We note that under more restricted cinematical conditions 
(and thus with a weaker result) a similar analysis
was performed in \cite{FSBY}. 
We can resume our findings in the following
\begin{lemma}
\label{syle}
The amplitude $\,\Ga_{\mu \nu\rho}(p_1,p_2,p_3)\,$ can be written 
in the form (\ref{expr}) where the scalar amplitudes $A$, $B$ depend 
on the euclidean invariants $p_1^2,\ p_2^2,\ p_3^2\,$ only.\\ 
If  $\,\Ga_{\mu \nu\rho}(p_1,p_2,p_3)\,$ is analytic at vanishing momentum, 
as  is the case in a  fully massive
theory, its  dependence on $A$ (\ref{expr})
excludes any  relevant local contribution to  
 $\,\Ga_{\mu \nu\rho}(p_1,p_2,p_3)\,$. 
\end{lemma}
Note that {\it analyticity  thus would exclude  the appearance of 
an  anomaly}\footnote{this conclusion is based on the complete
Bose symmetry of the amplitude. In more complicated theories
like the standard model there are fermion triangle contributions
which are not fully Bose symmetric due to the presence of 
several vector boson species. In this case the previous conclusion 
does not hold.}. 
We also note that a local contribution to  
$\,\Ga_{\mu \nu\rho}(p_1,p_2,p_3)\,$ compatible with the symmetries 
has to be at least of dimension 6. The corresponding term 
in the lagrangian then takes the form
\[
(\pa_\rho A_\rho )\, \ti F^{\mu \nu} \,F_{\mu \nu}\ ,\quad
\mbox{with}\quad
\ti F^{\mu \nu} \equiv \vep_{\mu \nu\rho \si}\, F^{\rho \si}\ .
\] 
The results of Lemma \ref{syle} are confirmed at
1-loop order by explicit calculation in \ref{PVCalc}.

\subsection{Explicit results on the Pauli-Villars regularized 1-loop amplitude}
\label{PVCalc}

We consider the one-loop triangular diagram (fig.\ref{Tux}), 
with complete Bose symmetry 
between the external legs.
In our case, we take an 
IR-regulator $\Lambda$.  We use a Pauli-Villars 
regularization\footnote{we hardly found any calculations
in the literature which are not based on the dimensional 
scheme. Sometimes  a global PV-regularization,
obtained on introducing a heavy fermion
is used \cite{Bert}. If not directly applied to the integrand,
this  still does not lead to well-defined integrals, however.} 
of the fermionic propagators:
\eq
\frac{\slashed{k}}{k^2}\ \rightarrow\ \frac{\slashed{k}}{k^2}
\ \sigma^F_{\Lambda,\Lambda_0}(k)
=S^{\LLz}  (k)\ ,
\eqe
and we introduce Feynman parameters with the following notations:
\eq
\int d\mu_n \,\equiv\, \left(\prod_{i=1}^n \int_0^{1} dx_i\right)
\delta(1-\sum_{i=1}^n x_i)\ , 
\eqe
\eq  
x_{i_1...i_m}\,=\, x_{i_1}+\dots + x_{i_m}\ ,\quad 
\overline{x}_{i_1...i_m}\,=\, 1-x_{i_1...i_m}\ , \quad 
\tilde{x}_{i_1...i_m}\,=\, 
x_{i_1...i_m}\ \overline{x}_{i_1...i_m} \ .
\eqe

\begin{figure}
\begin{center}
\centerline{\mbox{\epsfysize 4cm \epsffile{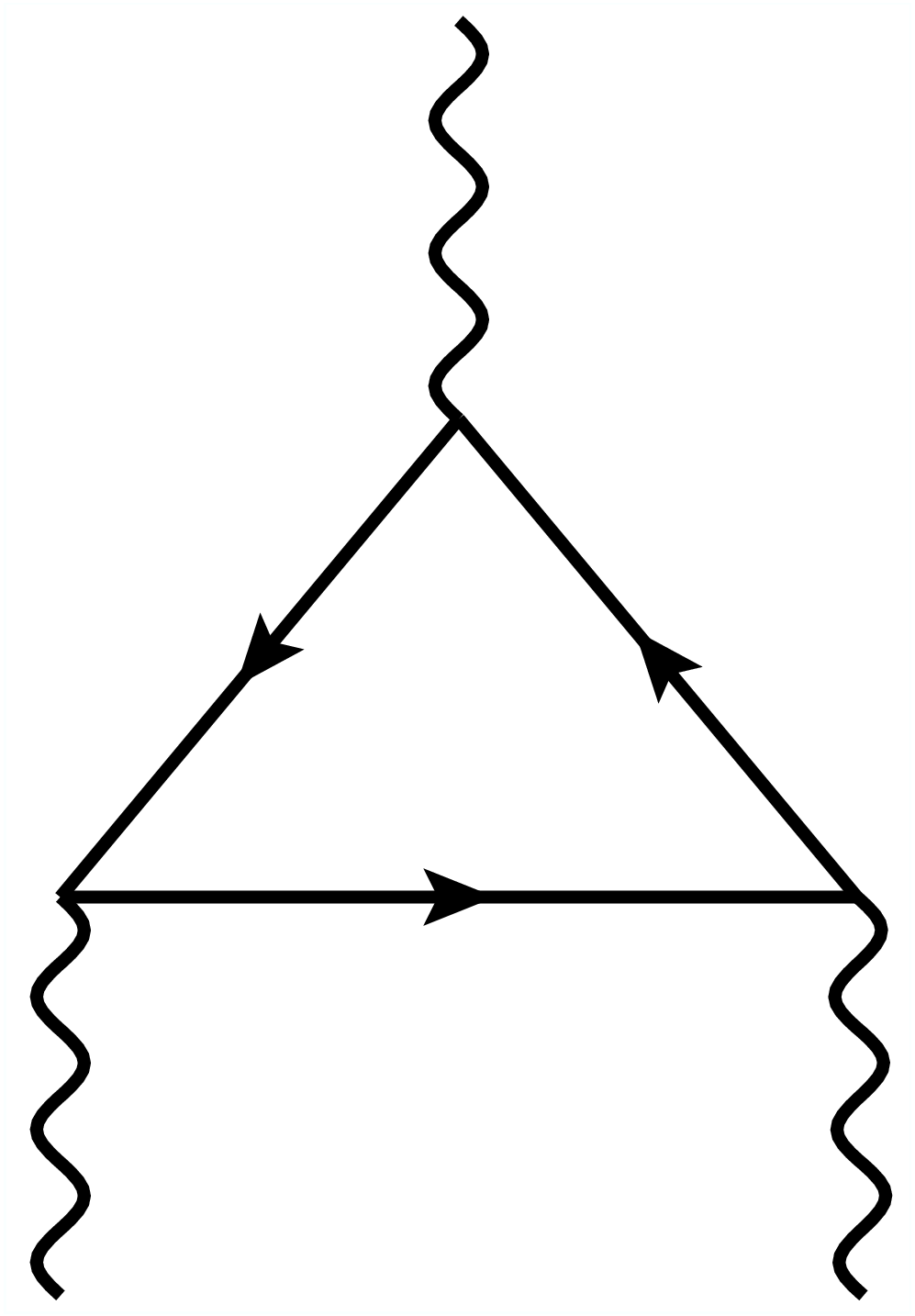}}}
\caption{The triangular diagram}
\label{Tux}
\end{center}
\end{figure}

\begin{enumerate}

\item

\textbf{The regularized (symmetrized) 
one-loop triangle} diagram is given by
\begin{eqnarray*}
\Gamma_{\mu\nu\rho}^{\LLz}=
2\int_k \textbf{tr} \left[ \gamma_5  S^{\LLz}  (k)\gamma_\nu
  S^{\LLz}  (k-p_2) 
\gamma_\mu S^{\LLz}  (k+p_3)  \gamma_\rho   \right]\ ,
\end{eqnarray*}
and at one-loop order, the amplitudes $A$ and $B$ introduced previously 
are given by:
\begin{eqnarray*}
A(1,2,3)&=&\frac{\alpha^3}{\pi^2} \int \frac{d\mu_6 }{D^3}
\left( 2x_{56}-x_{1234}\right) \\
&+&\frac{\alpha^3}{\pi^2}\int \frac{d\mu_6 }{D^4}\Bigl[
p_1^2\left[
  x_{34}^2(1-x_{34}-3x_{12})-x_{56}^2(1-x_{56})+x_{34}(x_{12}-x_{56})  
\right]\\
&&\qquad\qquad
 +\, p_2^2\left[ 3x_{56}(x_{12}^2+x_{34}^2)+2x_{12}x_{34}-x_{56}x_{1234}    
\right]\\
&&\qquad\qquad
 +\, p_3^2\left[ x_{12}^2(1-x_{12}-3x_{34})-x^2_{56}(1-x_{56})
+x_{12}(x_{34}-x_{56}) \right]\Bigr]
\end{eqnarray*}
and
\begin{eqnarray*}
&&B(1,2,3)=\frac{2\alpha^3}{\pi^2}\int \frac{d\mu_6 }{D^4}
\left[x_{34}(x_{12}^2+x_{56}^2)
  -3x_{12}x_{34}-x_{56}(x_{12}^2+x_{34}^2)  
\right]\ ,
\end{eqnarray*}
with
\[
D\equiv D(\Lambda,\Lao;p_1,p_2,p_3;x_1,\ldots,x_6)
\equiv x_{135}\Lambda^2+x_{246}\Lambda_0^2+
\tilde{x}_{34}p_2^2+\tilde{x}_{56}p_3^2+2x_{34} x_{56}p_2\cdot p_3\ .
\]

$\Gamma_{\mu\nu\rho}^{\LLz}$ stays finite in the IR and UV limits
i.e. 
finite when we take first $\Lambda\rightarrow 0$ and then
 $\Lambda_0\rightarrow \infty\,$.

\item 
\textbf{The contracted triangle}
For $\Lambda=0$ we obtain
\begin{eqnarray}
p_{1\mu} \Gamma_{\mu\nu\rho}^{0,\Lambda_0}=\frac{2}{\pi^2} 
\epsilon_{\nu\rho \alpha\beta}  p_{2\alpha}p_{3\beta} \int d\mu_5
\frac{x_3 \Lambda_0^6}{\left[\tilde{x}_{25}p_2^2
+\tilde{x}_3p_3^2+2x_{25}x_3p_2\cdot p_3+x_{123}\Lambda_0^2\right]^3}
\end{eqnarray} 
and in the UV limit, only one integral
\begin{eqnarray*}
\lim_{\Lambda_0\rightarrow 
\infty}\lim_{\Lambda\rightarrow 0}
\left(p_{1\mu}\Gamma_{\mu\nu\rho}^{\Lambda,\Lambda_0}\right)
=\frac{2}{\pi^2}\epsilon_{\nu\rho\alpha\beta }p_{2\alpha}p_{3\beta}
\int d\mu_6 \frac{x_3}{x_{123}^3}
\end{eqnarray*}
survives. We find explicitly
\begin{eqnarray}\label{anom1}
\lim_{\Lambda_0\rightarrow 
\infty}\lim_{\Lambda\rightarrow 0}
\left(p_{1\mu}\Gamma_{\mu\nu\rho}^{\Lambda,\Lambda_0}\right)
=\frac{1}{6\pi^2}\,\epsilon_{\nu\rho\alpha\beta }\,p_{2\alpha}\,p_{3\beta}\ .
\end{eqnarray}

\item 
\textbf{Derivatives of the triangle} The first momentum derivatives are 
finite for $\Lambda_0\rightarrow 
\infty,\Lambda\rightarrow 0\,$. 
In the second 
derivatives logarithmic divergences show up 
for exceptional momentum configurations 
when $\Lambda\rightarrow 0\,$. For example, we find 
\eq
\left.\delta_{\alpha\beta}\frac{\partial^2
 \Gamma^{\Lambda,\Lambda_0}_{\mu\nu\rho}}{\partial p_{2\alpha}
\partial p_{2\beta}}\right|_{p_2=0,p_3\neq 0}^{(div)}
=\ \frac{1}{2\pi^2}\epsilon_{\mu\nu\rho\sigma}\,
\frac{p_{3\sigma}}{p_3^2+\Lambda^2}\,
\ln\left(\frac{\Lambda^2}{\mu ^2}\right)\ ,
\eqe
where $\mu ^2>0$ is an arbitrary momentum scale. 
Here the superscript is justified by the fact that 
\eq
\left.\delta_{\alpha\beta}\frac{\partial^2
 \Gamma^{\Lambda,\Lambda_0}_{\mu\nu\rho}}{\partial p_{2\alpha}
\partial p_{2\beta}}\right|_{p_2=0,p_3\neq 0}
-\ 
\left.\delta_{\alpha\beta}\frac{\partial^2
 \Gamma^{\Lambda,\Lambda_0}_{\mu\nu\rho}}{\partial p_{2\alpha}
\partial p_{2\beta}}\right|_{p_2=0,p_3\neq 0}^{(div)}
=\,p_{3\sigma}\,\epsilon_{\mu\nu\rho\sigma}\ f(\Lambda,\mu,p_3^2)
\eqe
with 
\[
|f(\Lambda,\mu ,p_3^2)|\,\leq \, 
\frac{20}{\pi^2\sqrt{p_3^2(p_3^2+4\Lambda^2)}}
\ln\left( 1+\sqrt{\frac{p_3^2+4\Lambda^2}{p_3^2}}  \right)\ .
\]
 Thus the r.h.s. is finite for all values of 
$\Lambda\,$ if $\,p_3\neq 0\,$.
\end{enumerate}

\section{On the Jacobian of regularized
  BRS-transformations \label{appa}}
In this appendix we prove Lemma \ref{jac}. 
We consider a finite cube of side length $L$  in $\mr^4\,$,
and we expand the field variables in terms of plane waves 
\[
\Bigl\{ e^{ik_nx} \ \Bigl|\ k_n=(k_{n_0},k_{n_1},
k_{n_2},k_{n_3}),\ k_{n_i}=\frac{2\pi n_i}{L},\ n_i\in \mathbb{Z}   \Bigr\}
\]
thus imposing periodic boundary conditions
\footnote{Our result also holds if we take antiperiodic boundary 
conditions for fermions and/or ghosts.}
\begin{eqnarray*}
&&A_\mu(x)=\sum_{n\in \mathbb{Z}^4}A_{\mu,n}\,e^{ik_nx}\ , \quad
\psi^i(x)=\sum_{n\in \mathbb{Z}^4}\psi_{i,n}\,e^{ik_nx}\ ,
\quad
\overline{\psi}^i(x)=\sum_{n\in \mathbb{Z}^4}\overline{\psi}_{i,n}\,e^{ik_nx}
\ ,\\
&&c(x)=\sum_{n\in \mathbb{Z}^4}c_{n}\,e^{ik_nx}\ ,
\quad
\overline{c}(x)=\sum_{n\in \mathbb{Z}^4}\overline{c}_{n}\,e^{ik_nx}\ .
\\
\end{eqnarray*}In the regularized theory we want to calculate the 
Jacobian $J$ of the BRS transform of the
integration measure for all field modes. This measure  
 can be written as 
\begin{eqnarray*}
\left( \prod_{\mu=0}^3\prod_{n\in \mathbb{Z}^4}dA_{\mu,n}\right)\
\left(\prod_{j=1}^4\prod_{n\in \mathbb{Z}^4}d\psi_{j,n} 
\,d\overline{\psi}_{j,n}\right)\
\left( \prod_{n\in \mathbb{Z}^4}dc_n \,
d\overline{c}_n \right) \,\equiv \, 
\prod_{i=1}^{14}\prod_{n\in \mathbb{Z}^4}d\phi_{i,n} \ .
\end{eqnarray*} 
We 
write $\Phi(x)=\left\{  \phi_i(x),i=1,...,14\right\}$ for the set of 
all components of the fields of the theory with 
$A_\mu=\phi_{\mu+1},\ \psi^i=\phi_{4+i},\ \overline{\psi}i
=\phi_{8+i},\ c=\phi_{13},\ \overline{c}=\phi_{14}\,$.\\  
$J\,$ is the determinant of a matrix 
$ \left(\left(M_{ji}\right)_{n'n}\right)$ 
built of blocks $14\times 14$, with indices taking values in $\mathbb{Z}^4$
\begin{eqnarray*}
\left(M_{ji}\right)_{n'n}=\frac{\partial \phi'_{i,n}}{\partial
  \phi_{j,n'}}\ .
\end{eqnarray*}
We will call  the matrix 
of elements $\frac{\partial \phi'_{i,n}}{\partial \phi_{j,n'}}\,$ 
with $\,i,j\in [1,14]\,$, the $(n,n')$-block of $M\,$.\\
The regularized BRS-transformations then induce the following
changes of the field variables:
\begin{eqnarray*}
 \sum_{n\in \mathbb{Z}^4}A'_{\mu,n}e^{ik_nx}
&=&\sum_{n\in \mathbb{Z}^4}A_{\mu,n}e^{ik_nx}- 
\sum_{n\in \mathbb{Z}^4} R_1^0
\int dy\, \sigma_{0,\Lambda_0}(x-y)ik_{n,\mu} c_n\, e^{ik_ny}\,\epsilon \\
\Rightarrow\  A_{\mu,n}'&=&A_{\mu,n}
- i\, R_1^0\,\sigma_{0,\Lambda_0}(k_n)\,k_{n,\mu} \, c_n\,\epsilon\ ,\\
 \sum_{n\in \mathbb{Z}^4} \psi'_{i,n}\,e^{ik_nx}
&=&\sum_{n\in \mathbb{Z}^4}\psi_{i,n}\,e^{ik_nx}
-i\,g\, R_2^0\,\sum_{n,m\in \mathbb{Z}^4} 
\int dy\, \sigma_{0,\Lambda_0}(x-y)\,\psi_{i+2,n}\,c_m\,
e^{i(k_n+k_m)y}\,\epsilon\\ 
\Rightarrow\ \psi'_{i,n}&=&\psi_{i,n}
-i\,g\, R_2^0\, \sigma_{0,\Lambda_0}(k_n)
\sum_{n_1+n_2=n}\psi_{i+2,n_1}c_{n_2}\, \epsilon \ ,
\end{eqnarray*}
and similarly
\[
\overline{\psi}'_{i,n}
=\overline{\psi}_{i,n}-i\,g\,R_3^0\,\sigma_{0,\Lambda_0}(k_n)
\sum_{n_1+n_2=n}\overline{\psi}_{i+2,n_1}c_{n_2}\, \epsilon\ , \\
\]
\[
 c'_n=c_n\ , \quad
\overline{c}'_n
= \overline{c}_n-\frac{i\,R_4^0}{\alpha}\,
\sigma_{0,\Lambda_0}(k_n)\, k_{n}\,A_{n}\, \epsilon \ .
\]
We first study the  diagonal blocks for which $n=n'\,$, 
and then  the non-diagonal ones.

\begin{enumerate}
\item \textbf{Diagonal blocks} We obtain
\begin{eqnarray*}
&&\frac{\partial A'_{\mu,n}}{\partial A_{\nu,n}}= \delta_{\mu\nu}
\ , \ \quad 
\frac{\partial A'_{\mu,n}}{\partial c_n}=
-i\,R_1^0\,k_{n,\mu}\,\sigma_{0,\Lambda_0}(k_n)\,\epsilon\\
&&\frac{\partial \psi'_{i,n}}{\partial \psi_{j,n}}
=\delta_{ij}+i\,g\,R_2^0\,\sigma_{0,\Lambda_0}(k_n) 
\delta_{i+2,j}c_{(0,0,0,0)}\,\epsilon 
\ , \quad 
\frac{\partial \psi'_{i,n}}{\partial c_n}
=i\,g\,R^0_2\,\sigma_{0,\Lambda_0}(k_n)\,\psi_{i+2,(0,0,0,0)}\,
\epsilon\ , \\
&&\frac{\partial \overline{\psi}'_{i,n}}{\partial\overline{\psi}_{j,n}}
=\delta_{ij}+i\,g\,R_3^0\,\sigma_{0,\Lambda_0}(k_n) 
\delta_{i+2,j} c_{(0,0,0,0)}\,\epsilon
\ , \quad
\frac{\partial\overline{\psi}'_{i,n}}{\partial c_n}
= i\,g\,R_3^0\, \sigma_{0,\Lambda_0}(k_n)\,
\overline{\psi}_{i+2,(0,0,0,0)}\,\epsilon\ , \\
&&\frac{\partial c^{\,'}_n}{\partial c_ n}=1\ , \ \quad
\frac{\partial\overline{c}^{\,'}_n}{\partial \overline{c}_n}
=1\ , \ \quad
\frac{\partial \overline{c}^{\,'}_n}{\partial A_{\mu,n}}
=-i\,\frac{R_4^0}{\alpha}\,\sigma_{0,\Lambda_0}(k_n)\,k_{n,\mu}\,\epsilon \ .\\
\end{eqnarray*}
All other coefficents are zero.

\item \textbf{Non-diagonal blocks $(n,n')\,$}. We have the relations
\begin{eqnarray*}
\frac{\partial\psi'_{i,n}}{\partial \psi_{j,n'}}
=\delta_{i+2,j}i\,g\,R_2^0\,\sigma_{0,\Lambda_0}(k_n)\, c_{n-n'}\,\epsilon
\ , \quad 
\frac{\partial\psi'_{i,n}}{\partial c_{n'}}=i\,g\,R_2^0\,
\sigma_{0,\Lambda_0}(k_n)\, \psi_{i+2,n-n'}\,\epsilon\\
\frac{\partial\overline{\psi}'_{i,n}}{\partial \overline{\psi}_{j,n'}}
=\delta_{i+2,j}i\,g\,R_3^0\,\sigma_{0,\Lambda_0}(k_n) \,c_{n-n'}\,\epsilon
\ , \quad
\frac{\partial\overline{\psi}'_{i,n}}{\partial c_{n'}}
=i\,g\,R_3^0\,\sigma_{0,\Lambda_0}(k_n)\,
\overline{\psi}_{i+2,n-n'}\,\epsilon\ .\\
\end{eqnarray*}
All other elements of this block are zero. 
\end{enumerate}

We then deduce an explicit expression for a general block $(n,n')$:
\begin{eqnarray*}
M_{n'n}=\delta_{n,n'}\begin{pmatrix}
\mathbf{1}_4 & 0 & 0 & 0 & M_1\\
0& \mathbf{1}_4 & 0 & 0 & 0 \\
0 & 0 & \mathbf{1}_4 & 0 & 0 \\
M_2 & 0 & 0 & 1 & 0\\
0 & 0 & 0 & 0 & 1\\
\end{pmatrix}
+
\begin{pmatrix}
0 & 0 & 0 & 0 & 0 \\
0 & M_3 & 0 & 0 & 0\\
0 & 0 & M_4 & 0 & 0\\
0& M_5 & M_6 & 0 & 0\\
0 & 0 & 0  &0 & 0\\
\end{pmatrix}
\end{eqnarray*}
with
\begin{eqnarray*}
M_1&=&    -i\,\frac{R_4^0}{\alpha}\,\sigma_{0,\Lambda_0}(k_n)\epsilon \ , \\
M_2&=&  -i\,R_1^0\,\sigma_{0,\Lambda_0}(k_n)\begin{pmatrix}
k_{n,0} & k_{n,1}& k_{n,2} & k_{n,3}
\end{pmatrix}\,\epsilon\ ,  \\
M_{3/4}&=&  i\,g\,R_{2/3}^0\,\sigma_{0,\Lambda_0}(k_n)\, c_{n-n'}
\begin{pmatrix}
0 & 0 & 1 & 0\\
0&0 & 0 & 1 \\
1&0&0&0 \\
  0 & 1&0&0 \\
\end{pmatrix}  \epsilon \ ,  \\
M_5&=&i\,g\,R_2^0\,\sigma_{0,\Lambda_0}(k_n)
\begin{pmatrix}
\psi_{3,n-n'}&\psi_{4,n-n'}&\psi_{1,n-n'}&\psi_{2,n-n'}\\
\end{pmatrix}\,\epsilon  \ ,    \\
M_6&=&i\,g\,R_3^0\,\sigma_{0,\Lambda_0}(k_n)
\begin{pmatrix}
\overline{\psi}_{3,n-n'}&\overline{\psi}_{4,n-n'}&
\overline{\psi}_{1,n-n'}&\overline{\psi}_{2,n-n'}\\
\end{pmatrix}\,\epsilon  \ .   \\
\end{eqnarray*}
To obtain a well-defined finite dimensional determinant we introduce 
an UV cutoff $\Lao$ through restricting the sum over Fourier modes
to $|k_n| \le \Lao\,$
\[
\phi_{i,reg}(x)=\sum_{n\in\mathbb{Z}^4,\, |k_n| \le \Lao} 
\phi_{i,n}\,e^{ik_nx}\ ,
\]
as stated in Lemma \ref{jac}.
Due to the cutoff the matrix $\left(\frac{\partial 
\phi'_{i,n}}{\partial \phi_{j,n'}}\right)$ is finite-dimensional, and 
we can apply the usual formula for the determinant of an  
$\,n\times n$-matrix 
$M$, i.e.
$\,
det(M)=\sum_{\sigma\in S_n}\epsilon(\sigma)\prod_{i=1}^n M_{i,\sigma(i)}
\,$.
A nonvanishing  element $\alpha=(M_{ij})_{nn'}\neq 0,1\,$ 
 is  of order $\epsilon\,$. Consider a contribution
$A\,$ to the determinant for which  $\alpha$ contributes.  On the same
line as $\alpha$, in the $(n,n)$-block, there is a unique nonvanishing
coefficient $\beta$ which equals 1.
 $A$ is a multiple of a coefficient of the column of $\beta$
(different from $\beta$). But apart from $\beta$, the only non-zero
elements in this column are of order $\epsilon\,$. 
Then $A$ is zero because  $\epsilon^2=0\,$.\qed

\section{Comments on Fujikawa's argument  \label{fuji}} 

Fujikawa's argument \cite{Fu}, \cite{FuS} links the chiral anomaly to the
appearance of a Jacobian in the BRS transformation of the functional 
measure of integration. The argument is reproduced in many textbooks.
From the mathematical point of view there are loopholes in this
argument which we try to put into evidence, and it seems that
the interpretation of Fujikawa's calculation, in particular
in which sense precisely it may be related to the chiral anomaly, 
is unclear.\\
The arguments proceeds from a decomposition of the fermionic fields 
w.r.t.  eigenbases 
$\left\{\phi_n\right\},\left\{\phi_n^\dagger\right\}$
\begin{eqnarray*}
\psi(x)&=&\sum_n a_n \phi_n(x)  \quad \Rightarrow \quad \psi'(x)
\equiv e^{i\alpha(x) \gamma_5}\,\psi(x)
=\sum_n a'_n \phi_n(x)\ ,\\
\overline{\psi}(x)&=&\sum_n b_n \phi^\dagger_n(x)  \quad  
\Rightarrow \quad
\overline{\psi}^{\,'}(x) \equiv \overline{\psi}(x)\,e^{i\alpha(x) \gamma_5}
=\sum_n b'_n \overline{\phi}^\dagger_n(x)\ ,
\end{eqnarray*}
with
\begin{eqnarray*}
a'_m&=&a_m+i\sum_n a_n \int dx\, \alpha(x) 
\phi_m^\dagger(x)\gamma_5\phi_n(x)\ ,\\
b'_m&=&b_m+i\sum_n b_n \int dx\, \alpha(x)
\phi_n^\dagger(x)\gamma_5\phi_m(x)\ .
\end{eqnarray*}
Thus the Jacobian of this transformation is
\begin{equation*}
\det\left(1+i\int dx\, \alpha(x)\,\phi^\dagger_n(x)\, \gamma_5\, \phi_m(x)   
\right)^{-2}\ ,
\end{equation*}
because the variables $a_n,\ b_n$ are Grassmannian. Using the matrix 
relation $\det(M)=\exp\{Tr\ln(M)\}$ with 
$\,M=1\,+\,i\int dx\, \alpha(x)\,\phi^\dagger_n(x)\,\gamma_5\, \phi_m(x)\,$, 
and expanding  the logarithm to first  order in $\alpha$, we obtain 
the Jacobian
\begin{equation}\label{Jack1}
\prod_n\exp\left(-2i\int dx\, \alpha(x)\,\phi_n^\dagger(x)\,
\gamma_5\,\phi_n(x)    \right)\ .
\end{equation}
Using the plane wave basis of section \ref{appa} for the fermionic modes
(on introducing cutoffs), we would conclude that this jacobian equals 1.

Fujikawa regularizes (\ref{Jack1}) in a way depending on the  
vector field which is viewed as a background field. In fact he
introduces a smooth function $f(x)\,$
such that 
$f(0)=1,\ f(\infty)=0\,$, and writes a regularized Jacobian:
\begin{equation*}
\prod_n\exp\left(-2i\int dx\, 
\alpha(x)\ \phi_n^\dagger(x)\ \gamma_5\ 
f(\frac{\slashed{D}^2}{M^2})\ \phi_n(x)    
\right).
\end{equation*}
Here   $\slashed{D}= \gamma_\mu \pa_\mu - i e A_\mu\,$
is the covariant Dirac operator, and $M < \infty\,$ 
is an UV regulator. The functions
$\phi_n$ are then supposed to be eigenfunctions
of  $\slashed{D}\,$. In this case the previous expression
is well-defined only if the spectrum of  $\slashed{D}\,$ is discrete
which generically will not be the 
case\footnote{a {\it necessary} condition would be that the field
$A_\mu(x)\,$ diverges for $|x| \to \infty\,$.}.
In the next step one passes to a plane wave basis using the relation
\begin{equation}
\lim_{M\rightarrow \infty} \sum_{n=1}^\infty 
\int dx\ \alpha(x)\ \phi_n^\dagger(x) 
\gamma_5 \ f( \frac{\slashed{D}^2}{M^2})\ \phi_n(x)
=
\lim_{M\rightarrow \infty}\textbf{tr}\int dx\ 
\alpha(x) \int_k e^{-ikx}\gamma_5\ 
f(\frac{\slashed{D}^2}{M^2}) \ e^{ikx}  \label{eq1}\ ,
\end{equation}
where $\textbf{tr}$ indicates 
the spinor space trace. 
Applying the operator $f\left(\frac{\slashed{D}^2}{M^2}\right)\,$ to
$e^{ikx}$, 
and performing the change of variables $\,k_\mu\rightarrow Mk_\mu\,$, 
(\ref{eq1}) becomes
\eq
\label{eq2}
\exp\left( -2i \lim_{M\rightarrow \infty}M^4\textbf{tr} 
\int dx\,\alpha(x)\int_k \gamma_5 f \Bigl( (ik_\mu+\frac{D_\mu}{M})^2
-\frac{ie}{4}\left[\gamma^\mu,\gamma^\nu\right]\frac{F_{\mu\nu}}{M^2}      
\Bigr)\right)\ ,
\eqe
since $\slashed{D}^2=-D^2-\frac{ie}{4}
\left[\gamma^\mu,\gamma^\nu\right]F_{\mu\nu}\,$. 
Expanding $f$ around $k_\mu$ and observing  that only terms of 
order $\, \geq  -4\, $ in $\, M\, $ and containing at least $4\, $  
$\gamma$-matrices, survive for $\, M \to \infty\,$, (\ref{eq2}) becomes
\begin{eqnarray}
\label{ff}
&&\exp\left(\frac{ie^2}{8}\textbf{Tr}\left(\gamma_5\left[\gamma^\mu,
\gamma^\nu\right]\left[\gamma^\rho,\gamma^\sigma\right]  \right)\int
dx\,
\alpha(x)F_{\mu\nu }(x)F_{\rho\sigma}(x)  \int_k  f\left( k^2
\right) 
\right)\\
\nonumber
&=&\exp\left( 2ie^2 K  \int dx\,
 \alpha(x)\tilde{F}^{\mu\nu}(x) F_{\mu\nu}(x)\right)\ ,
\end{eqnarray}
where $K\,$ is a constant {\it depending on the function $\,f\,$.} 
Choosing  $f(x)=e^{-x}$ one finds $K=\frac{1}{16\pi^2}$.
The  contribution of this Jacobian then gives rise to an
anomalous term in the divergence of the axial current
of the form
\begin{eqnarray*}
p_\mu  \langle \,j_5^\mu(p)\,\rangle \, =\ 2i\,m \, \langle \,
j_5(p)\, 
\rangle 
\, + \ \frac{ie^2}{8\pi^2}\,
\langle \,  (\tilde{F}_{\mu\nu }F^{\mu\nu})(p)\, \rangle \qquad
\mbox{(for fermions of mass } \ m) \ .
\end{eqnarray*}
The result (\ref{ff}) has been obtained by introducing a 
background field dependent regulator for the fermions.
Regularizing the fermions modes independently of this 
background would produce a trivial Jacobian as shown in App. \ref{appa}. 
The mathematical questions raised previously could be 
circumvented saying that what has been calculated is 
the short-time limit of the trace 
involving the diagonal part of the heat kernel $K_t(x,y)\,$
of the operator 
\[
\exp\{-t \slashed{D}^{\,2}\}
\]
in such a background field \cite{Lue}:
\begin{equation}
\lim_{t \to 0} \, \textbf{tr}  \int dx\, \alpha(x)\,  \gamma_5 \, K_t(x,x)\ .
\end{equation}
But it is then not clear why this quantity should be directly
related to the Jacobian of the chiral  gauge transformation of the
fermion fields.


\begin{thebibliography}{99}

\bibitem{KM} Ch. Kopper, V.F. M\"uller: Renormalization of
  spontaneously broken SU(2) Yang-Mills theory with flow equations,
Rev. Math. Phys. {\bf 21}, 781-820 (2009).

\bibitem{Ad1} S.L. Adler: Axial-Vector vertex in spinor
  electrodynamics, Phys. Rev. {\bf 177}, 2426-2438 (1968).

\bibitem{Ad2} S.L. Adler, W.A. Bardeen: Absence of higher-order 
corrections in the anomalous axial-vector divergence equation, 
Phys. Rev {\bf 182}, 1517-1536 (1969).

\bibitem{BJ} J.S. Bell, R. Jackiw, A PCAC puzzle: 
$\pi ^{0} \to \ga \ga\,$ in the $\si$-Model,
Nuovo Cim.{\bf A 60}, 47-61 (1969) 

\bibitem{Bert} R.A. Bertlmann: \emph{Anomalies in quantum field
    theory}, Clarendon Press, Oxford (1996). 

\bibitem{Bilal} A. Bilal: Lectures on anomalies, 
arXiv: hep-th/0802.0634 (2008).

\bibitem{BIM} C. Bouchiat, J. Iliopoulos, Ph. Meyer: An anomaly-free version
 of Weinberg's model, Phys. Letters, {\bf 38B}, 519-523 (1972).

\bibitem{BRS}  C. Becchi, A. Rouet and R. Stora: Renormalization of
        gauge theories, \\ Ann. Phys. (N.Y.) {\bf 98}, 287 - 321 (1976). 

\bibitem{Fa}  P. Falco: Vector and axial anomaly in the Thirring-Wess model,
J. Math. Phys. {\bf 51} 082306  (2010).


\bibitem{FS} L.D. Faddeev, A.A. Slavnov, \emph{Gauge fields: introduction
             to quantum theory}, 2nd edition\  Addison-Wesley Pub.,
Reading MA (1991). 

\bibitem{FP} 
V.N.  Popov and L.D. Faddeev~  Kiev Inst. Theor. Phys. Acad. 
Sci. preprint ITP 67-36 (1967), English translation: Perturbation
 theory for gauge-invariant fields (preprint NAL-THY-57, 1972). 
Reprinted e.g. in \emph{50 years of Yang-Mills theory}, ed. G 't Hooft, 
World Scientific (2005).


\bibitem{FSBY} Y. Frishman, A. Schwimmer, T. Banks, S. Yankielowicz: 
The axial anomaly and the bound-state spectrum in confining theories, 
Nucl. Phys. {\bf     B177}, 157-171 (1981).

\bibitem{Fu} K. Fujikawa: Path integral measure for gauge-invariant 
theories, Phys. Rev. Lett. {\bf 42}, 1195-1198 (1979).

\bibitem{FuS} K. Fujikawa, H. Suzuki:  
\underline{Path integral and quantum anomalies}, 
Oxford University Press (2004).

\bibitem{GJ} D.J. Gross, R. Jackiw: Effect of anomalies in
  quasi-renormalizable theories, Phys. Rev. D {\bf 6}, 
477-493 (1972).
  
\bibitem{GK} R. Guida, Ch. Kopper: Uniform momentum bounds via flow equations:
            Massless $\varphi_4^4$, to appear.

\bibitem{KK} G. Keller, Ch. Kopper~: Perturbative renormalization of composite
operators via flow equations I. Commun. Math. Phys. {\bf 148}, 
445-467 (1992).  

\bibitem{KKQ}  G. Keller, Ch. Kopper: 
Perturbative renormalization of QED via flow equations,
Phys. Lett. {\bf B273}, 323-332 (1992).

\bibitem{KK94} 
G. Keller, Ch. Kopper: Perturbative renormalization of massless
         phi**4 in four dimensions with flow equations. 
Commun. Math. Phys. {\bf 161} 515-532 (1994).

\bibitem{KKQ2}  G. Keller, Ch. Kopper: Renormalizability 
proof for QED based on flow equations, 
Commun. Math. Phys. {\bf 176}, 193-226 (1996). 

\bibitem{KKS} G. Keller, Ch. Kopper, M. Salmhofer: 
Perturbative renormalization
and effective Lagrangians in $\Phi^4_4$, Helv. Phys. Ac\-ta {\bf 65}, 32-52
(1991).

\bibitem{Le} B. L\'ev\^eque: Thesis Ecole Polytechnique, to appear.

\bibitem{Lue} M. L\"uscher: Dimensional regularization in the presence
  of large background fields, Ann. Phys. (N.Y.) {\bf 142}, 359-392
(1982).

\bibitem{Ma}  V. Mastropietro: Non perturbative Adler-Bardeen theorem, \\
 J. Math. Phys. {\bf 48}, 022302 (2007).

\bibitem{M} V.F. M\"uller:  Perturbative renormalization by flow 
    equations,\  Rev. Math. Phys. {\bf 15}, 491 - 558 (2003),
arXiv:hep-th/0208211. 

\bibitem{Po} J. Polchinski: Renormalization and effective Lagrangians,\
Nucl.Phys. {\bf B231}, 269-295 (1984).

\bibitem{Ty} I.V. Tyutin: Gauge invariance in field theory and statistical
      mechanics, \
  Lebedev FIAN 39 (1975).   

\bibitem{WH} F. Wegner, A. Houghton: Renormalization group equations
  for critical phenomena, Phys. Rev. {\bf  A8}, 401-412 (1973). 

\bibitem{Wi}  K. Wilson: Renormalization group and critical phenomena 
I. Renormalization group and the Kadanoff scaling picture,
Phys.Rev. {\bf B4}, 3174-3183 (1971),\\ 
K. Wilson: Renormalization group and critical phenomena
II. Phase cell analysis of
critical behaviour, Phys.Rev. {\bf B4}, 3184-3205  (1971).

\bibitem{Zi} J. Zinn-Justin:  \emph{Quantum field theory and critical
phenomena}, ch. 21, Clarendon Press,Ox\-ford, 3rd ed. (1997), and 
J. Zinn-Justin in: Trends in elementary particle theory, Lecture Notes
in Physics {\bf 37}, 2 - 40, Springer-Verlag (1975). 

\bibitem{Zi2} J. Zinn-Justin: Chiral anomalies and topology, 
arXiv: hep-th/0201220 (2002).



\end{thebibliography}
\end{document}